\documentclass[prd,twocolumn,a4paper,superscriptaddress,floatfix]{revtex4}
\usepackage{bbm}
\usepackage[all]{xy}
\usepackage{amsmath}
\usepackage{amssymb}
\usepackage[pdftex]{graphicx}
\usepackage{epstopdf}

\def\be{\begin{equation}}
\def\ee{\end{equation}}
\newcommand{\bes}{\begin{subequations}}
\newcommand{\ees}{\end{subequations}}

\def\ben{\begin{eqnarray}}
\def\een{\end{eqnarray}}
\def\ba{\begin{array}}
\def\ea{\end{array}}

\def\bb{\bibitem}
\begin{document}
\newcommand{\half}{{\textstyle\frac{1}{2}}}
\allowdisplaybreaks[3]
\def\a{\alpha}
\def\b{\beta}
\def\g{\gamma}\def\G{\Gamma}
\def\d{\delta}\def\D{\Delta}
\def\ep{\epsilon}
\def\et{\eta}
\def\z{\zeta}
\def\t{\theta}\def\T{\Theta}
\def\l{\lambda}\def\L{\Lambda}
\def\m{\mu}
\def\f{\phi}\def\F{\Phi}
\def\n{\nu}
\def\p{\psi}\def\P{\Psi}
\def\r{\rho}
\def\s{\sigma}\def\S{\Sigma}
\def\ta{\tau}
\def\x{\chi}
\def\o{\omega}\def\O{\Omega}
\def\k{\kappa}
\def\pa {\partial}
\def\ov{\over}
\def\br{\\}
\def\ud{\underline}

\newcommand\lsim{\mathrel{\rlap{\lower4pt\hbox{\hskip1pt$\sim$}}
    \raise1pt\hbox{$<$}}}
\newcommand\gsim{\mathrel{\rlap{\lower4pt\hbox{\hskip1pt$\sim$}}
    \raise1pt\hbox{$>$}}}
\newcommand\esim{\mathrel{\rlap{\raise2pt\hbox{\hskip0pt$\sim$}}
    \lower1pt\hbox{$-$}}}

\title{Bifurcation and pattern changing with two real scalar fields}

\author{P.P. Avelino}
\email[Electronic address: ]{ppavelin@fc.up.pt}
\affiliation{Centro de F\'{\i}sica do Porto, Rua do Campo Alegre 687, 4169-007 Porto, Portugal}
\affiliation{Departamento de F\'{\i}sica da Faculdade de Ci\^encias
da Universidade do Porto, Rua do Campo Alegre 687, 4169-007 Porto, Portugal}
\author{D. Bazeia}
\email[Electronic address: ]{bazeia@fisica.ufpb.br}
\affiliation{Departamento de F\'{\i}sica, Universidade Federal da Para\'{\i}ba
58051-970 Jo\~ao Pessoa, Para\'{\i}ba, Brasil}
\author{R. Menezes}
\email[Electronic address: ]{rms@fisica.ufpb.br}
\affiliation{Centro de F\'{\i}sica do Porto, Rua do Campo Alegre 687, 4169-007 Porto, Portugal}
\affiliation{Departamento de F\'{\i}sica da Faculdade de Ci\^encias
da Universidade do Porto, Rua do Campo Alegre 687, 4169-007 Porto, Portugal}
\affiliation{Departamento de F\'{\i}sica, Universidade Federal da Para\'{\i}ba
58051-970 Jo\~ao Pessoa, Para\'{\i}ba, Brasil}
\author{J.C.R.E. Oliveira}
\email[Electronic address: ]{jespain@fe.up.pt}
\affiliation{Centro de F\'{\i}sica do Porto, Rua do Campo Alegre 687, 4169-007 Porto, Portugal}
\affiliation{Departamento de Engenharia F\'{\i}sica da Faculdade de Engenharia
da Universidade do Porto, Rua Dr. Roberto Frias, s/n, 4200-465 Porto, Portugal}

\begin{abstract}
This work deals with bifurcation and pattern changing in models described by two real scalar fields. We consider generic models with quartic potentials and show that the number of independent polynomial coefficients affecting the ratios between the various domain wall 
tensions can be reduced to $4$ if the model has a superpotential. We then study specific one-parameter families of models and compute the wall tensions associated with both BPS and non-BPS sectors. We show how bifurcation can be associated to modification of the patterns of domain wall networks and illustrate this with some examples which may be relevant to describe realistic situations of current interest in high energy physics. In particular, we discuss a simple solution 
to the cosmological domain wall problem.
\end{abstract} 
\maketitle

\section{Introduction}
 Bifurcation is a well known phenomenon, which occurs in several distinct areas of nonlinear science \cite{R1}.
Usually, it can be controlled by some forcing parameter, which drives the system to change from a given behavior to another. Interesting examples can be found, for instance, in a competition between the roll and hexagon structures which appear in an experiment of thermal convection \cite{C}, in nematic liquid crystals subject to rotating magnetic fields where spiral waves were observed \cite{LC}, and in periodically forced nonlinear oscillatory systems, where patterns are seen to spring induced by bifurcation, as reported for instance in \cite{E1,Fr1}.

A distinguishing feature of these studies is that they are all non-relativistic. However, bifurcation also appears in various high energy physics models where static domain wall solutions are shown to bifurcate under the variation of model parameters \cite{AT}. Other examples are given in \cite{O1}, relevant for the cosmological evolution in brane-world scenarios, in \cite{O2}, associated with critical gravitational collapse, and in \cite{D,O3}, in the context of hybrid inflation.

In the present work, we take a specific direction by studying bifurcation in a class of relativistic models described by two real scalar fields, $\phi$ and $\chi$. The basic motivation is to study the possibility that a network of domain walls may bifurcate into different configurations, giving rise to modifications of the network pattern. If we choose two minima arbitrarily, it may be possible that they are connected with two or more distinct orbits. When this happens the system develops a bifurcation, since one can go from a given minimum to another one, following two or more distinct paths. In one dimension, given three minima $j$, $k$, and $l$, the direct $jl$ path is energetically favored over the path $jk+kl$ if the energy density or tension of the $(jl)$ structure is lower than the sum of the other two tensions, that is $\tau_{jl}<\tau_{jk} +\tau_{kl}$. A bifurcation occurs when the inequality becomes an equality, since we can go from $j$ to $l$ following the direct $jl$ path, or alternatively visiting $k$ through the path $jk+kl$, at no extra 
energy cost. In more than one dimension the analysis needs to take into consideration the intersection angle of the domain walls.

In this paper we consider generic models with quartic potentials with or without a superpotential. In the former case the analysis is simpler 
since the tension of the domain walls associated with a given BPS sector is fully determined by the values of the 
superpotential, $W(\phi,\chi)$, at the minima of the potential. We take $W(\phi,\chi)$ to be a polynomial function of degree $3$ which gives rise to a quartic potential, $V(\phi,\chi)$ (note that for superpotentials of order lower than $3$ there are no domain wall solutions). 

The situation is more complicated when the two-field model is not the bosonic portion of some supersymmetric theory. In this case, there is no superpotential and the determination of the domain wall tensions is not so straightforward. However, an interesting  procedure was followed  in \cite{BBL} where a model (the so-called BBL model) with two real scalar fields, $\phi$ and $\chi$, was constructed under the assumption that its potential could be written as the sum of two quartic potentials, $V(\phi)$ and $V(\chi)$, plus a small interaction term.  Although the model as a whole has no superpotential, the importance of the interaction term can be controlled by adjusting a small parameter, $\varepsilon$. The domain wall tensions could then be determined analytically for $|\varepsilon| \ll 1$.

The evolution of the macroscopic properties of domain wall networks is strongly dependent on whether or not junctions are present and, in the former case, on the dimensionality of the junctions. Although the presence of junctions is not enough to make domain walls a viable dark energy candidate, their impact on the dynamics of the network needs to be taken into account in any quantitative study of cosmological consequences of domain wall evolution \cite{ammmo1}. In this paper we perform domain wall network simulations of a new class of one-parameter multi-tension domain wall models, showing that it has a number of interesting properties. In particular, we discuss a simple solution to the cosmological domain wall problem which arises naturally in this class of models.

In this paper we study bifurcation and pattern changing in models described by two real scalar fields, with or without a superpotential. The outline of this paper is as follows. We start by reviewing the BBL model in Section II, analytically and numerically calculating  the domain wall tensions and comparing our results with analytical results, valid for $|\varepsilon| \ll 1$, obtained in  \cite{BBL} 
(see also  \cite{ammmo}). 
We discuss how bifurcation arises in this model as well as the corresponding implications for the dimensionality of the junctions 
which may be present when we consider more than one spatial dimension. In section III we study the other possibility, that is, we consider models which support bifurcation and are described by a superpotential. We investigate this case in detail because it is expected to be of more general interest, since it can be seen as the bosonic portion of a larger, supersymmetric theory. We also 
discuss some implications of our findings in the light of the cosmological domain wall problem.
We end the paper in section IV with some comments and conclusions.

\section{Absence of superpotential}
Let us start by reviewing the two field model introduced in \cite{BBL}. Consider two real scalar fields $\phi$ and $\chi$ with the Lagrangian
\be
{\cal L}=\frac12\partial_\mu\phi\partial^\mu\phi+\frac12\partial_\mu\chi\partial^\mu\chi-V(\phi,\chi)\,,
\ee
where the potential $V$ is given
\be
V(\phi,\chi)=V(\phi)+V(\chi)+\varepsilon u(\phi,\chi)\,,
\ee
with $\varepsilon$ being a small parameter, which control the interaction between the two fields,
\be
V(\phi)=\frac12 \left(r-\frac{\phi^2}{r}\right)^2,\;\;\;\;\;V(\chi)=\frac12 \left(r-\frac{\chi^2}{r}\right)^2\,,
\ee
\be
u(\phi,\chi)=\frac14(\phi^4+\chi^4-6\phi^2\chi^2+4r^4)\,,
\ee
and $r$ is a real number. The numerical factors are introduced for convenience. We are considering dimensionless variables, that is, we have rescaled both the fields and the space and time coordinates appropriately such that they are now dimensionless quantities.

If $\varepsilon=0$ then the fields, $\phi$ and $\chi$, are not coupled and the minima of the potential are at $v^0_i=(\bar\phi_i,\bar\chi_i)$ with $v^0_1=(-r,r)$, $v^0_2=(-r,-r)$, $v^0_3=(r,-r)$ and $v^0_4=(r,r)$. For $\varepsilon$ very small the minima move slightly to new positions $v_i^{\varepsilon}$. In this limit the tensions can be calculated as \cite{abl} 
\bes\ben
\tau_e^\varepsilon=\tau_e^0+\varepsilon \int_{-\infty}^{\infty} dx u(\phi_e^0,\chi_e^0)\,, \label{tensionel}
\\
\tau_d^\varepsilon=\tau_d^0+\varepsilon \int_{-\infty}^{\infty} dx u(\phi_d^0,\chi_d^0)\,,
\een\ees
where $\tau_e^0=4r^2/3 $, $\tau^0_d=8 r^2/3$ are the tensions for the edge and the diagonal solutions  $\phi_e^0$ and $\phi_d^0$, respectively 
(with similar expressions for the $\chi$-field). The exact solutions with $\varepsilon=0$ can thus be used to obtain the tensions 
$\tau^\varepsilon_e=4r^2(1+7\varepsilon r^2/4)/3$ and $\tau^\varepsilon_d=8r^2(1+\varepsilon r^2)/3$ (see Refs.~\cite{BBL,ammmo} for more details). 

\begin{figure}[t!]
\includegraphics[width=9.0cm]{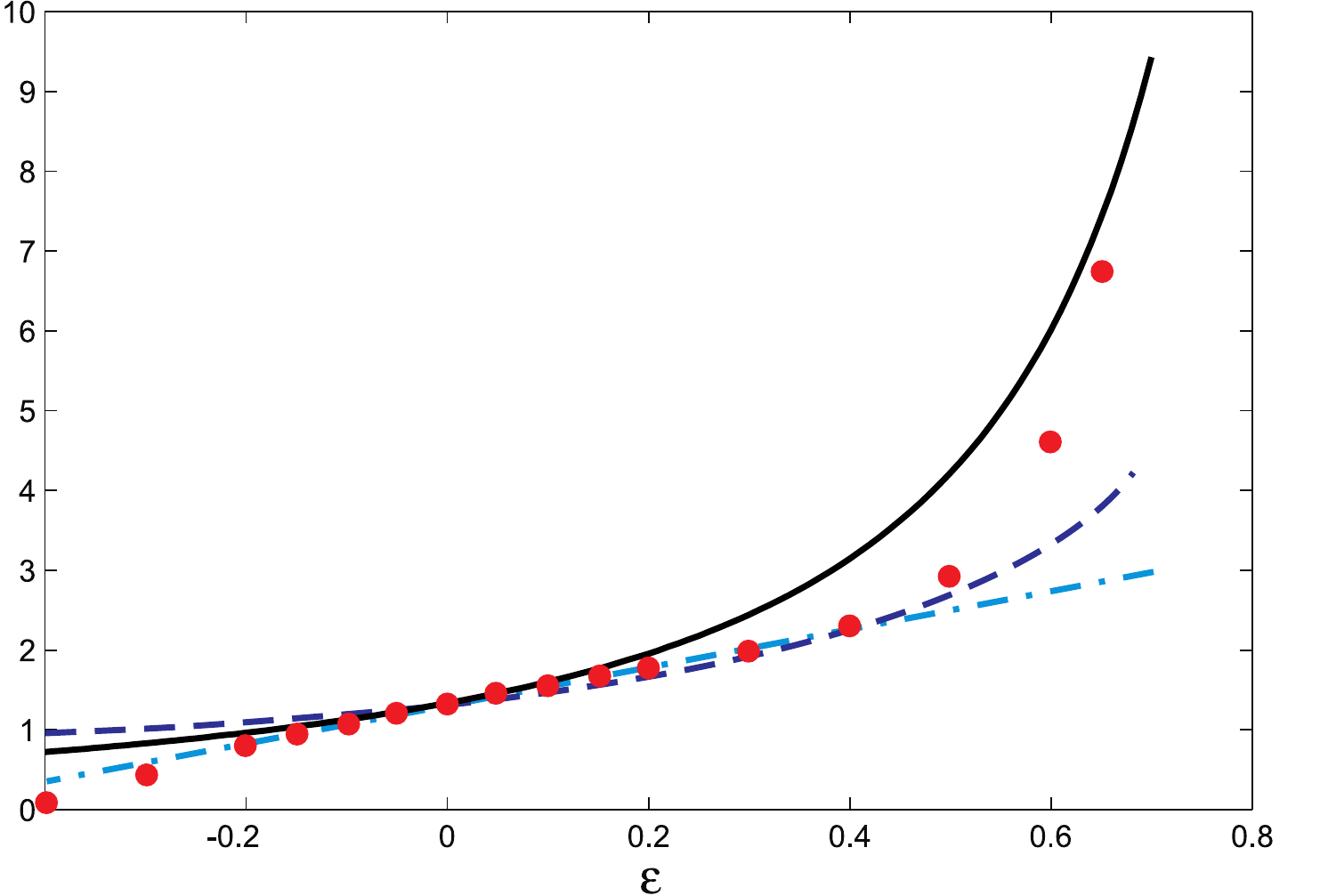}
\caption{\small{Comparison of the values of $\tau_e$ determined analytically, using either the straight line approximation  (curved solid line) or the linear approximation (straight dot-dashed line), with the numerical solution (filled circles), as a function of $\varepsilon$. The curved dashed line represents the value of $\tau_d/2$ calculated using Eqn. (\ref{tensiond}). We have 
taken $r=1$.}}
\end{figure}

By considering straight line orbits connecting two distinct minima
at the edge and diagonal sectors, we get the corresponding tensions
\bes\ben
\tau_e^{sl}&=&\frac43 r^2 \sqrt\frac{1+\varepsilon r^2/2}{(1-\varepsilon r^2)^3}\, \label{tensione}
\\
\tau_d^{sl}&=&\frac{8r^2}{3(1-\varepsilon r^2)}\,. \label{tensiond}
\een\ees
In fact, due to the symmetry of the scalar field potential, the straight diagonal orbit is a solution of the equations of motion, so that $\tau_d=\tau_d^{sl}$ corresponds to an exact 
solution. Although the edge solutions are not straight lines, Eqn. (\ref{tensione}) provides an approximation to the real tension up to first order in $\varepsilon$. We have also computed $\tau_e$ by solving the equations of motion for $\phi$ and $\chi$ numerically (see Ref. \cite{ammmo1} for details about the numerical code).

In Fig. 1 we compare the values of $\tau_e$ determined numerically as a function of $\varepsilon$ (filled circles), 
with the analytical expectations for $|\varepsilon| \ll 1$ determined using either the straight line approximation (Eqn. (\ref{tensione}) - curved solid line) or the linear approximation (Eqn. (\ref{tensionel}) - straight dot-dashed line). The curved dashed line represents the value of $\tau_d/2$ calculated using Eqn. (\ref{tensiond}). Here  we made the choice $r=1$. We see that the analytical results for $\tau_e$ are a good approximation of the exact 
solution for small values of $|\varepsilon|$,  becoming significantly less accurate for $|\varepsilon| \gsim 0.2$. We also see that 
$\tau_d/\tau_e$ is greater or smaller than $2$ depending on whether $\varepsilon$  is smaller or greater than zero, respectively. 
To investigate bifurcation, in one spatial dimension, we have to compare the diagonal tension with the sum of two edge tensions, and so we see that for $\varepsilon<0$ the model relaxes to the edge solutions, and for $\varepsilon>0$ it relaxes to the diagonal configuration. Thus, the sign of $\varepsilon$ induces bifurcation.

\begin{figure}[t!]
\hspace{-0.3cm}\includegraphics[width=4.5cm]{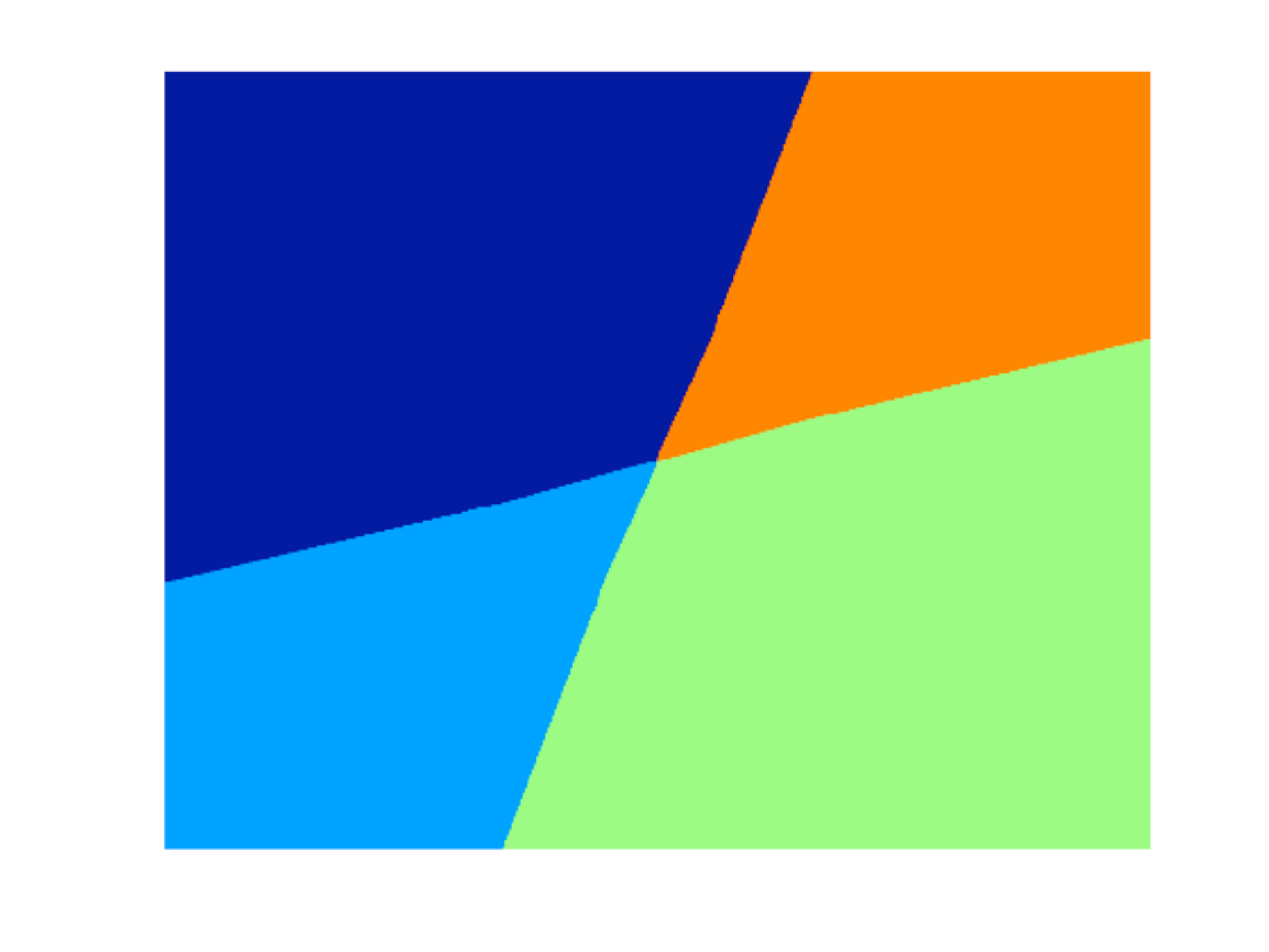}\hspace{-0.3cm}\includegraphics[width=4.5cm]{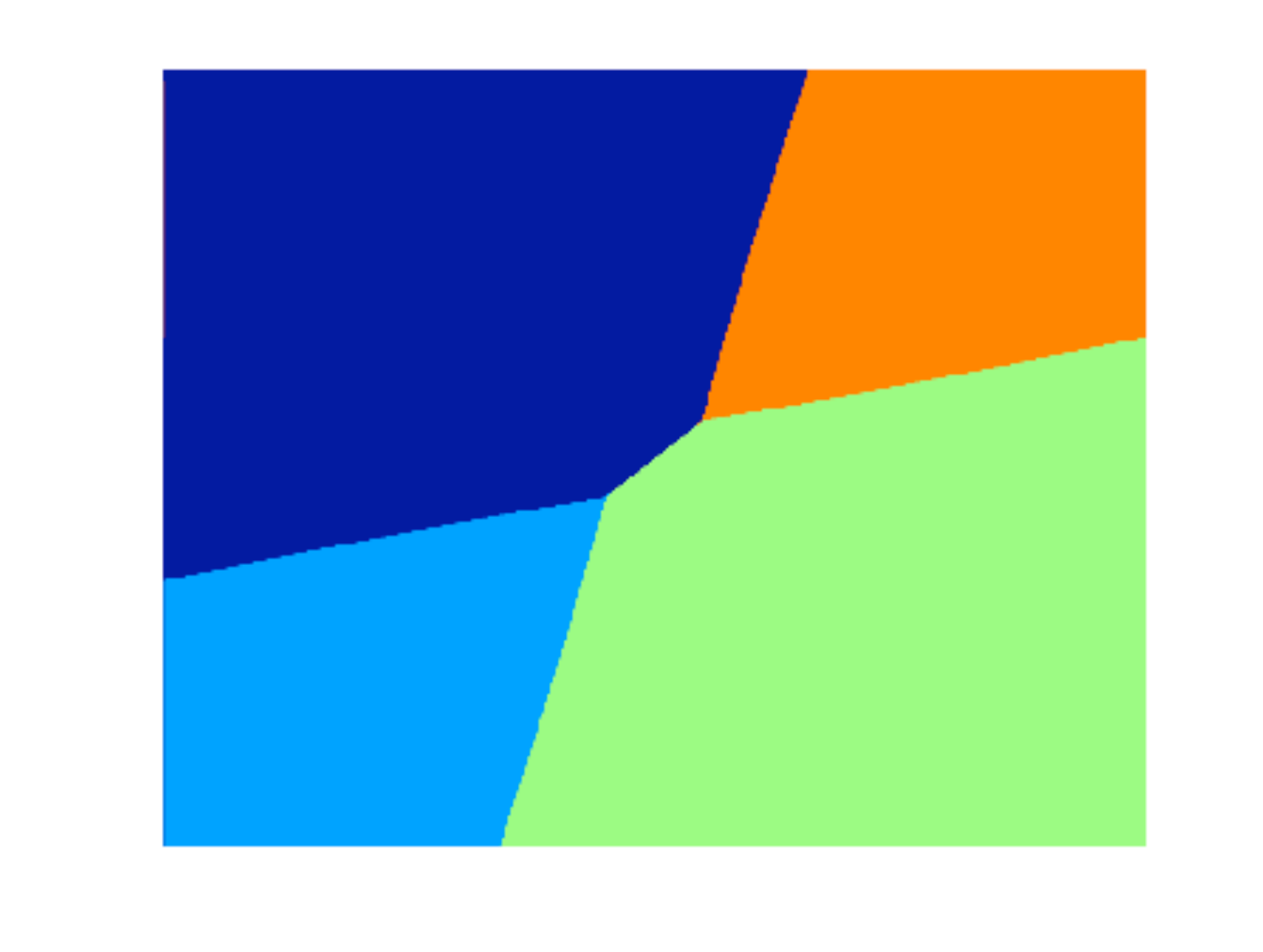}
\caption{\small{The left panel illustrates that for $\varepsilon < 0$ the model forms stable $X$-type junctions. The right panel shows that the formation of $Y$-type junctions is possible for $\varepsilon > 0$. We took $\varepsilon=-0.2$ and $\varepsilon=0.2$ in 
the numerical simulations illustrated in the left and right panels respectively.}}
\end{figure}

The above discussion in one dimension is crucial to determine whether $Y$-type or $X$-type junctions are favored in more 
than one dimension. If $\tau_d/\tau_e > 2$ ($\tau_d/\tau_e < {\sqrt 2}$) then only $X$-type ($Y$-type junctions) will be formed. 
There is an intermediate case corresponding to ${\sqrt 2} < \tau_d/\tau_e < 2$ in which both stable $Y$-type and $X$-type junctions are possible (depending on the intersection angle between the walls). 
However, if one starts with a $X$-type junctions where four walls intersect at right angles, it will be stable for $\tau_d/\tau_e > {\sqrt 2}$ and unstable for $\tau_d/\tau_e < {\sqrt 2}$. 
Let us consider the intersection angles $\theta_i$ between the walls that surround  each domain with minimum $v_i$ 
($i=1,2,3$ or $4$). Notice that $\theta_1=\theta_3$ and $\theta_2=\theta_4$. 
For $\varepsilon <0$ ($X$-type junctions) equilibrium is possible 
even if $\theta_1$ is different from $\theta_2$. Otherwise the domain walls would 
intersect at right angles which does not need to be the case, as it is clearly shown in the left panel of Fig.~2. For $\varepsilon > 0$ 
stable $X$-type junctions can still exist if $2 \tau_e \cos (\theta_2/2) < \tau_d$ (assuming that $\theta_2 \le \theta_1$). In this case, $\theta_2=\pi/2$  is the maximum possible value of the intersection angle. Consequently, we see that stable $X$-type junctions are also possible for any ${\sqrt 2} < \tau_d/\tau_e \le 2$, as long as 
$\theta_2 > 2 \, {\rm arcos}(\tau_d/(2\tau_e))$. For smaller values of $\theta_2$, $Y$-type junctions will form.

We illustrate some of these points in Fig.~2 where we plot the results of two different simulations of the BBL model in two spatial dimensions. We take $\varepsilon=-0.2$ and $\varepsilon=0.2$ in the left and right panels, respectively. In the first case, $\tau_d/\tau_e > 2$ and consequently $Y$-type junctions are unstable which leads to the formation of $X$-type junctions. In the second case, $\tau_d/\tau_e < 2$ and consequently the formation of $Y$-type is no longer forbidden. 
The multiplicity of the junctions is one of the most important variables affecting domain wall network  evolution and it is an essential aspect to consider in any quantitative assessment of the cosmological implications of domain wall networks \cite{ammmo1}.

\section{Presence of superpotential} 

In the present work we deal with bifurcation and pattern formation in two-field models, characterized by the real scalar fields $\phi$ and $\chi$, described by potentials containing up to fourth-order power terms. 

\subsection{Generic models with cubic superpotentials}

We shall now consider a model with two real fields characterized by the potential
\be
V(\phi,\chi)=\frac12\, W^2_\phi+\frac12\,W^2_\chi\,,  \label{vsuper}
\ee 
where $W=W(\phi,\chi)$ is the superpotential and
\be
W_\phi \equiv \frac{\partial W}{\partial\phi},\;\;\;\;\;W_\chi \equiv \frac{\partial W}{\partial\chi}\,.
\ee
The equations of motion for static solutions have the general structure
\bes\ben
\frac{d^2\phi}{dx^2}=W_\phi W_{\phi\phi}+W_\chi W_{\chi\phi}\,,
\\
\frac{d^2\chi}{dx^2}=W_\phi W_{\phi\chi}+W_\chi W_{\chi\chi}\,.
\een\ees
They are solved by solutions of the first-order equations
\bes\ben
\frac{d\phi}{dx}= \pm W_\phi\,,
\\
\frac{d\chi}{dx}=\pm W_\chi\,.
\een\ees
In the following we  shall drop the $\pm$ sign. It will be sufficient to realize that for each solution of the first order 
equations, $(\phi(x),\chi(x))$, there will also be another solution given by $(\phi(-x),\chi(-x))$.
The minima of the potential are the points in field space $(\bar\phi,\bar\chi)$ which solve the two algebraic equations $W_\phi=0$ and $W_\chi=0$, and so they are critical points of $W.$ We label the minima as $j,k,l,...,$ with $j=(\bar\phi_j,\bar\chi_j)$, etc. Each pair of distinct minima $j$ and $k$ $(k\neq j)$ forms a topological sector. A static solution is represented by $(\phi(x),\chi(x))$. If it  solves the first-order equations and connects two distinct minima in field space then this solution is a minimum energy configuration in this sector. It corresponds to topologically stable solutions, usually named BPS states in the supersymmetric model,  corresponding to a topological BPS sector -- see \cite{BB} for more details. A non-BPS sector is a topological sector which has no solution coming from solutions of the first-order equations. This is easy to identify because the tension associated with a BPS sector is given by $\tau_{jk}=|W(j)-W(k)| \neq 0$, and so a non-BPS sector is specified by the minima $l$ and $m$, say, 
which obey $W(l)=W(m)$.

We may write a generic potential of order $2(d-1)$ for the scalar fields $\phi$ and $\chi$ as 
\be
V(\phi,\chi)=\sum_{0 \le a+b \le 2(d-1)} C_{ab} \phi^a \chi^b\,.
\ee 
The number of coefficients, $C_{ab}$, of the polynomial potential is given by
\be
N_d=2 d^2 -d\,,
\ee
so that $N_2=6$ and $N_3=15$. On the other hand, if the above potential can be obtained from a polynomial superpotential of order $d$,
\be
W(\phi,\chi)=\sum_{1 \le a+b \le d} {\widetilde C}_{ab} \phi^a \chi^b\,,
\ee 
using Eqn.  (\ref{vsuper}), then the number of independent coefficients of the polynomial potential reduces to
\be
{\widetilde N}_d=\frac{d^2 +3d}{2}\,.
\ee
Here, $ {\widetilde C}_{ab}$ are the coefficients of the polynomial superpotential.
We see that ${\widetilde N}_2=5$ and ${\widetilde N}_3=9$. However, some of these are not real degrees of freedom in the 
sense that there are changes in the polynomial coeficients which do not affect the ratios between the domain wall tensions (assuming that the model has domain wall
solutions). 
These ratios are invariant under a change of the overall normalization of the potential $V \to \alpha_1 V$, a rescaling of the fields $(\phi, \chi) \to (\alpha_2 \phi, \alpha_2 \chi)$, a translation 
$(\phi,\chi) \to (\phi+\phi_*,\chi+\chi_*)$ or a rotation by an angle $\theta$ where $\alpha_1 > 0$, $\alpha_2$, $\phi_*$ 
and $\chi_*$ are real numbers. It is clear from the 
discussion that the case with $d=2$ is not interesting because the lowest polynomial order of $W$ 
for which one can obtain domain walls interpolating between different minima is $d=3$. In this case the initial number of 
coefficients, $9$, is reduced to $4$. In practice one can arbitrarily choose the positions of two of the minima at $v_1=(-1,0)$ and $v_3=(1,0)$ as well as the normalization of the potential and the remaining $4$  coefficients are associated with the positions of the other two minima $v_2=(x_2, y_2)$ and $v_4=(x_4, y_4)$.  

\subsection{A class of models with bifurcation} 

For simplicity we shall now consider a class of models with the two minima, $v_2$ and $v_4$, lying in the $yy$ axis so that $x_2=x_4=0$ and take $y_2=a$ and $y_4=b$. For $a \neq -b$  the most generic cubic superpotential satisfying all of the above conditions is
\be
W(\phi,\chi)=\lambda\left(-\frac{4ab}{a+b}+2\chi+\frac{4ab}{a+b}\phi^2-\frac43 \frac{1}{a+b} \chi^2\right)\chi\,.
\ee
If $a=-b$ the superpotential is given by
\ben
W(\phi,\chi)&=&\frac{\lambda_1}{3b^3} \left(-3 b^2 \chi+ 3 b^2 \phi^2 \chi + \chi^3\right) \nonumber\\
&+&\frac{\lambda_2}{3b^2} \left(-3 b^2 \phi + 3 \phi \chi^2 + b^2 \phi^3\right)\,.
\een
The standard BNRT model \cite{bsr} is obtained by making the choice $\lambda_1=0$ and $\lambda_2=-1$ but other choices of $\lambda_1$ and $\lambda_2$ also do not modify the positions of the minima $v_2=(0,-b)$ and $v_4=(0,b)$. 

In this paper we shall take $\lambda_1=b^2$ and $\lambda_2=0$ so that the superpotential becomes
\be
W(\phi,\chi)=-b \chi+b\phi^2 \chi+\frac{\chi^3}{3b}\,,
\ee
with $b > 0$. This particular choice is equivalent to a rotation of the original BNRT model by $\pi/2$ followed by a rescaling of the fields by a constant factor and consequently does not have any further implications. This model has a $Z_4$ symmetry if $b=1$ which is broken into $Z_2 \times Z_2$ for $b \neq 1$.

In this case the first order equations are given by
\bes\ben
\frac{d\phi}{dx}&=& 2 b \phi\chi \,,
\\
\frac{d\chi}{dx}&=&- b+\frac{\chi^2}{b}+b\phi^2\,.
\een\ees
The superpotential at the four minima is given by
\bes\ben
W(v_1)=0,\;\;\;\;\;W(v_2)=\frac23 b^2\,,
\\
W(v_3)=0,\;\;\;\;\;W(v_4)=-\frac23 b^2\,.
\een\ees
Thus, the model has one topological sector that is non-BPS, the sector $v_1\leftrightarrow v_3$, and five BPS sectors, one with tension $\tau_{24}=4 b^2/3$, and four with the same tension $\tau_{12}=\tau_{14}=\tau_{23}=\tau_{34}=2 b^2/3$. 
Also, the BPS sector $v_2\leftrightarrow v_4$ has an infinity of BPS solutions, degenerated to the same tension $\tau_{24}=4 b^2/3.$ A vertical straight line BPS solution is given by
\bes\ben
\phi(x)&=& 0\,,
\\
\chi(x)&=& -b\, {\rm tanh} (x)\,.
\een\ees 
The horizontal solution connecting the minima $v_1$ and $v_3$ is non-BPS and is given by
\bes\ben
\phi(x)&=&  {\rm tanh} (bx)\,,
\\
\chi(x)&=& 0\,,
\een\ees 
with tension $\tau_{13}^{sl}=4 b/3$.
This model has many interesting features. In particular, all the orbits can be found analytically in field space, as it was shown in Ref.~{\cite{igm}}. Another feature of this model is that it has $\tau_{24}=\tau_{12}+\tau_{14}$, independently of the value of $b$.
This has been studied in detail in Refs. \cite{BB,bsr,igm} and we shall now turn our attention to another one-parameter class of multi-tension domain wall models which has not appeared before in the literature. Unlike in the BNRT model, the non-diagonal sectors do not all have the same tension. We will see later that this property will be crucial in the light of a possible solution to the cosmological domain wall problem.

\begin{figure}[t!]
\includegraphics[width=7.0cm]{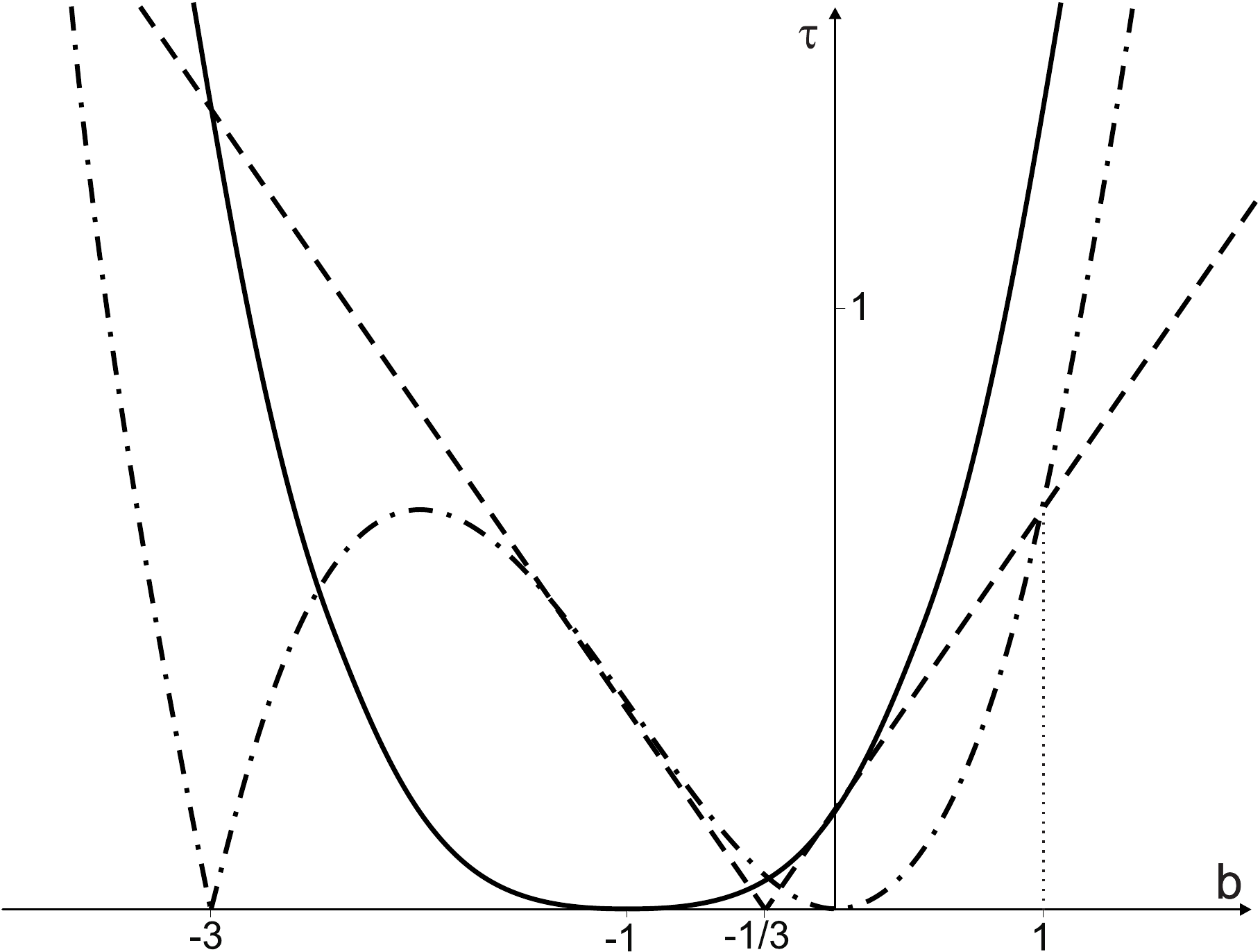}
\\
\vspace{0.3cm}
\includegraphics[width=7.0cm]{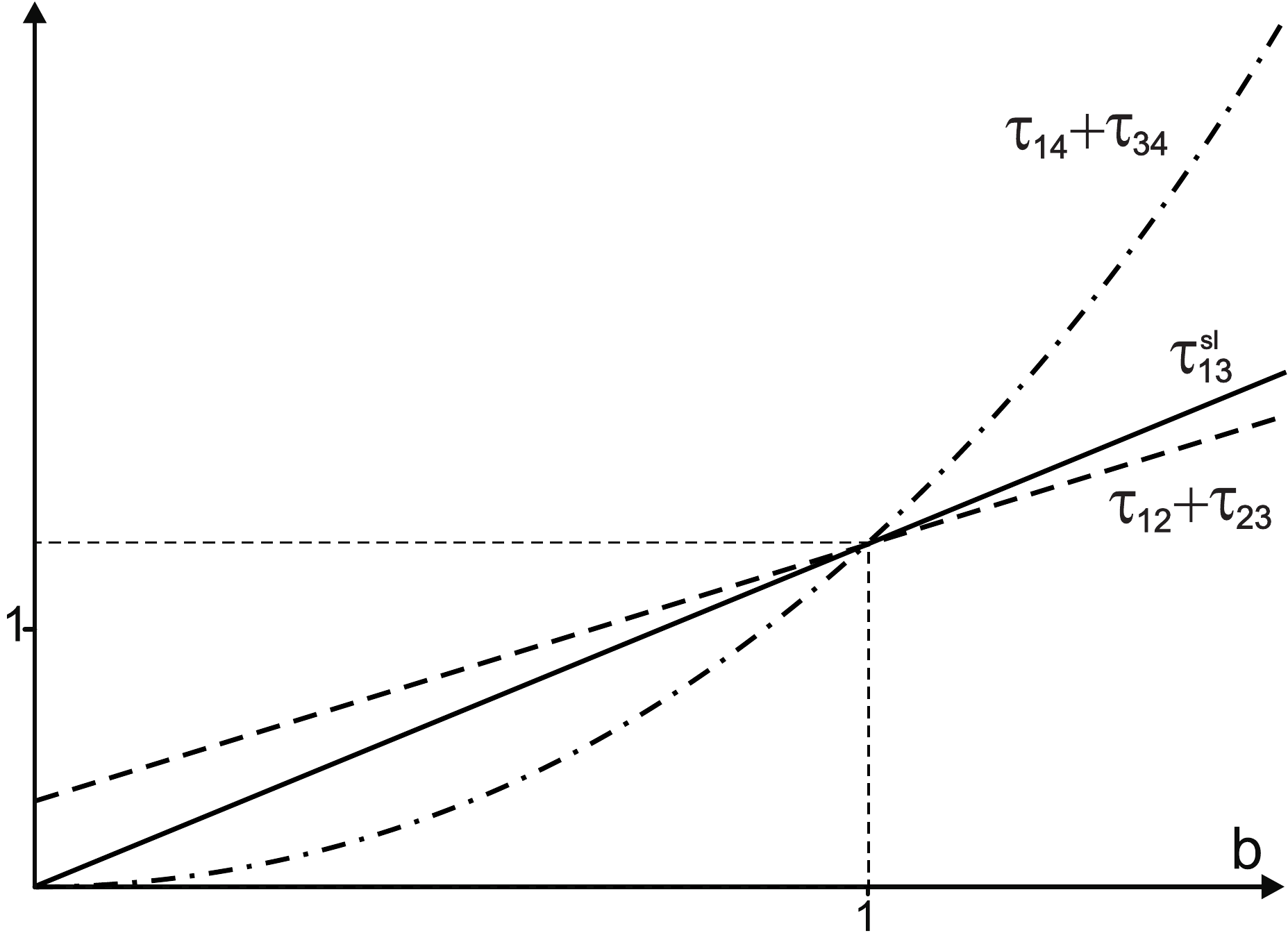}
\caption{\small{In the top panel we plot the BPS tensions as a function of $b$. Solid, dashed and dot-dashed lines represent the tensions $\tau_{24}$, $\tau_{12}$, and $\tau_{14}$, respectively. 
In the lower panel we plot the  sum of the BPS tensions $\tau_{14}+\tau_{34}$ 
and $\tau_{12}+\tau_{23}$ for $b >0$ as well as, $\tau_{13}^{sl}$, for the straight line non-BPS solution.  Note that if $b>0$ then 
$\tau_{24}=\tau_{12}+\tau_{14}$ and that $\tau_{14}=\tau_{34}$ and $\tau_{12}=\tau_{23}$ 
for any value of $b$. Also, notice that $\tau_{13}^{sl} \sim \tau_{13}$ if $b \sim 1$. }}
\end{figure}

The class of models we introduce in this paper is characterized by a cubic superpotential with minima at the values $v_1=(-1,0)$, $v_2=(0,-1)$, $v_3=(1,0)$ and $v_4=(0,b)$. For $b=1$ the model has a $Z_4$ symmetry which is broken into $Z_2$ for 
$b \neq 1$.

Apart from an arbitrary normalization, the positions of the minima completely 
determine the form of the superpotential which is given by 
\be
W(\phi,\chi)=\chi\left(-b + \frac12(1-b)\chi+b \phi^2+\frac13 \chi^2\right)\,, \label{superpot}
\ee
so that the potential is
\be
V(\phi,\chi)=\frac12\left((2b\phi\chi)^2+(-b+(1-b)\chi+b \phi^2 + \chi^2)^2\right)\, \label{potours}.
\ee
We see that if $b=0$ then $v_4$ is at the origin, that is $v_4=(0,0)$, and the minima $v_1$, $v_3$, and $v_4$ are aligned. When 
$b =1$ the four minima are at the vertices of a square.  Other special points are (see below): $b=-1/3$ giving $v_4=(0,-1/3)$, $b=-1$ making $v_4=(0,-1)$ coincident with $v_2$, and $b=-3$ giving $v_4=(0,-3)$.

In the case of a generic $b$, the first-order equations are
\bes\ben
\frac{d\phi}{dx}&=&2b\phi\chi\,, \label{phieq}
\\
\frac{d\chi}{dx}&=& -b+(1-b)\chi+b \phi^2 + \chi^2\,.\label{chieq}
\een\ees
There is a particular solution with $\phi=0$, which connects the minima $v_2$ and $v_4$, for which we have
\be
\frac{d\chi}{dx}=\left(\chi+\frac12 (1-b)\right)^2 -\frac14 (1+b)^2\,,
\ee
which has the explicit solution
\be\label{bps}
\chi(x)=-\frac12(1-b)-\frac12 (1+b)\tanh((1+b)x/2)\,.
\ee
Note that in the $b\to-1$ limit the minimum $v_4$ collapses into $v_2$, eliminating the topological sector $v_2\leftrightarrow v_4$.

The superpotential evaluated at the four minima gives
\bes\label{w}\ben
W(v_1)&=&0\,,\;\;\;W(v_2)=\frac16 \left(1+3b\right)\,,\\
W(v_3)&=&0\,,\;\;\;W(v_4)=-\frac16  b^2(3+b)\,,
\een\ees
and for $b=-1$ we get that $W(v_2)=W(v_4),$ in accordance with the above discussion. However, for any value of $b \neq-1$, the topological sector $v_2\leftrightarrow v_4$ is always a BPS sector, and the above particular solution \eqref{bps} is always a BPS solution. With \eqref{w}, we can then collect the tensions in all the BPS sectors. They are given by
\bes\ben
\tau_{12}&=&\tau_{23}=\frac16|1+3b|\,, \label{tau12}
\\
\tau_{24}&=&\frac16|b+1|^3\,, \label{tau24}
\\
\tau_{14}&=&\tau_{34}=\frac16 b^2|3+b|\,. \label{tau14}
\een\ees 
We plot the BPS tensions in the top panel of Fig.~3, to see how they behave as a function of $b$. Solid, dashed and dot-dashed lines represent the tensions $\tau_{24}$, $\tau_{12}$, and $\tau_{14}$, respectively. In particular, we note that $\tau_{24}=\tau_{12}+\tau_{14}$ for $b\leq-3$ and $b \geq -1/3$, $\tau_{14}=\tau_{12}+\tau_{24}$ for $-1 \le b \leq -1/3$, $\tau_{12}=\tau_{24}+\tau_{14}$ for $-3\leq b\leq-1$. 

For $b=0$ the three minima $v_1$, $v_3$, and $v_4$ get aligned, with $W(v_1)=W(v_3)=W(v_4)=0,$ so they cannot form BPS sectors. This case is uninteresting because we can make the $\phi$ field constant, with the $\chi$ field obeying 
\ben
\frac{d\chi}{dx}=(1+\chi)\chi\,,
\een
such that for $b=0$ the superpotential has the form
\be
W(\chi)=\frac12\chi^2+\frac13\chi^3\,.
\ee

The $b = 1$ case leads to a similar situation, and the fields $\phi_+=\phi+\chi$ and $\phi_-=\phi-\chi$ decouple, but now they both have nontrivial behavior. When $b=-1/3$, we get that $W(v_1)=W(v_2)=W(v_3)=0$ and $W(v_4)=-4/81$. In this case there are three BPS sectors with the same tension, $\tau_{14}=\tau_{24}=\tau_{34}=4/81$. For $b=-3$ we get $W(v_1)=W(v_3)=W(v_4)=0$ and $W(v_2)=4/3$.  Hence, in this case there are also three BPS sectors with the same tension, $\tau_{12}=\tau_{23}=\tau_{24}=4/3$. This is similar to the former case, for $b=-1/3$, with the minima $v_2$ and $v_4$ interchanged.

\begin{figure}[t!]
\includegraphics[width=3.5cm]{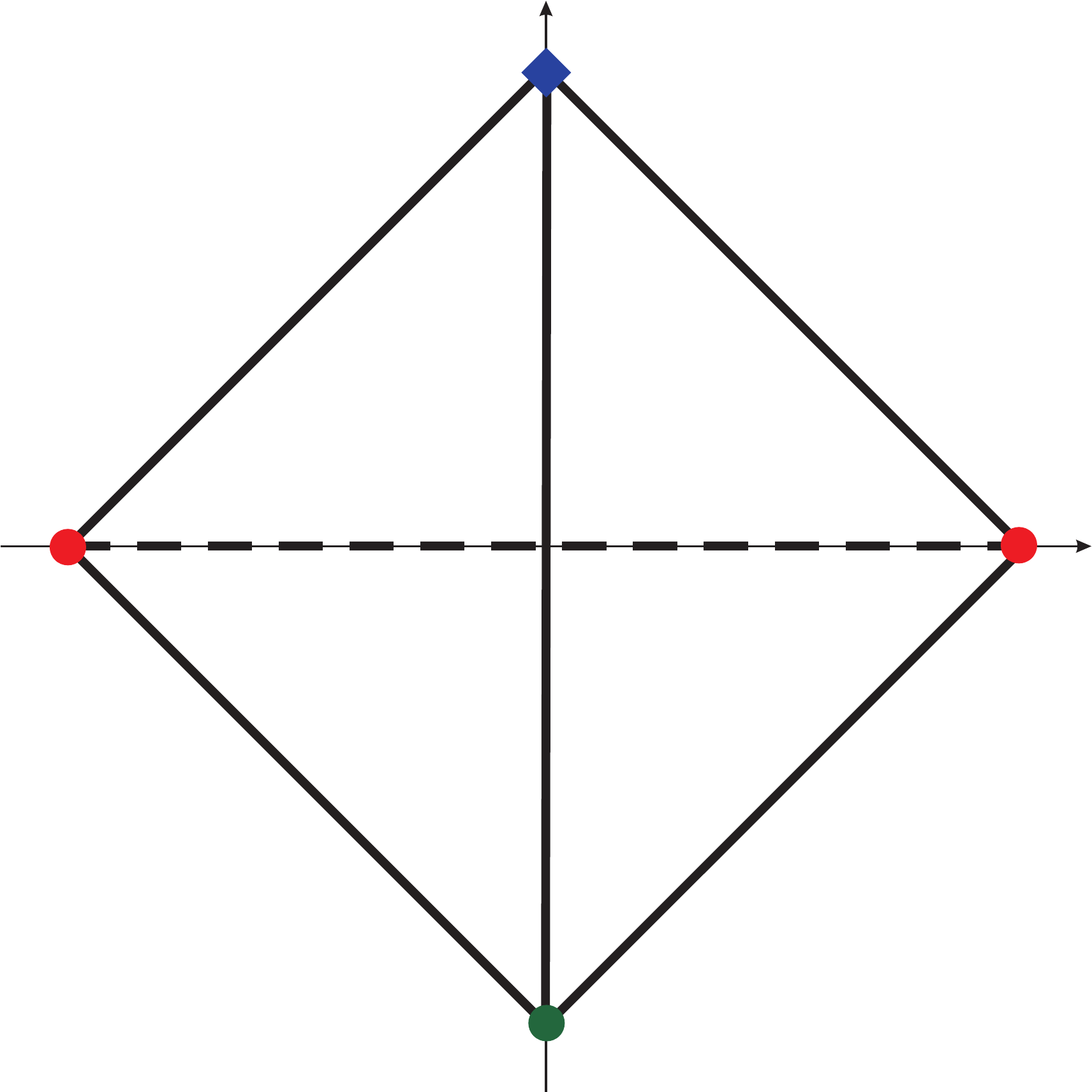}\hspace{0.3cm}\includegraphics[width=3.5cm]{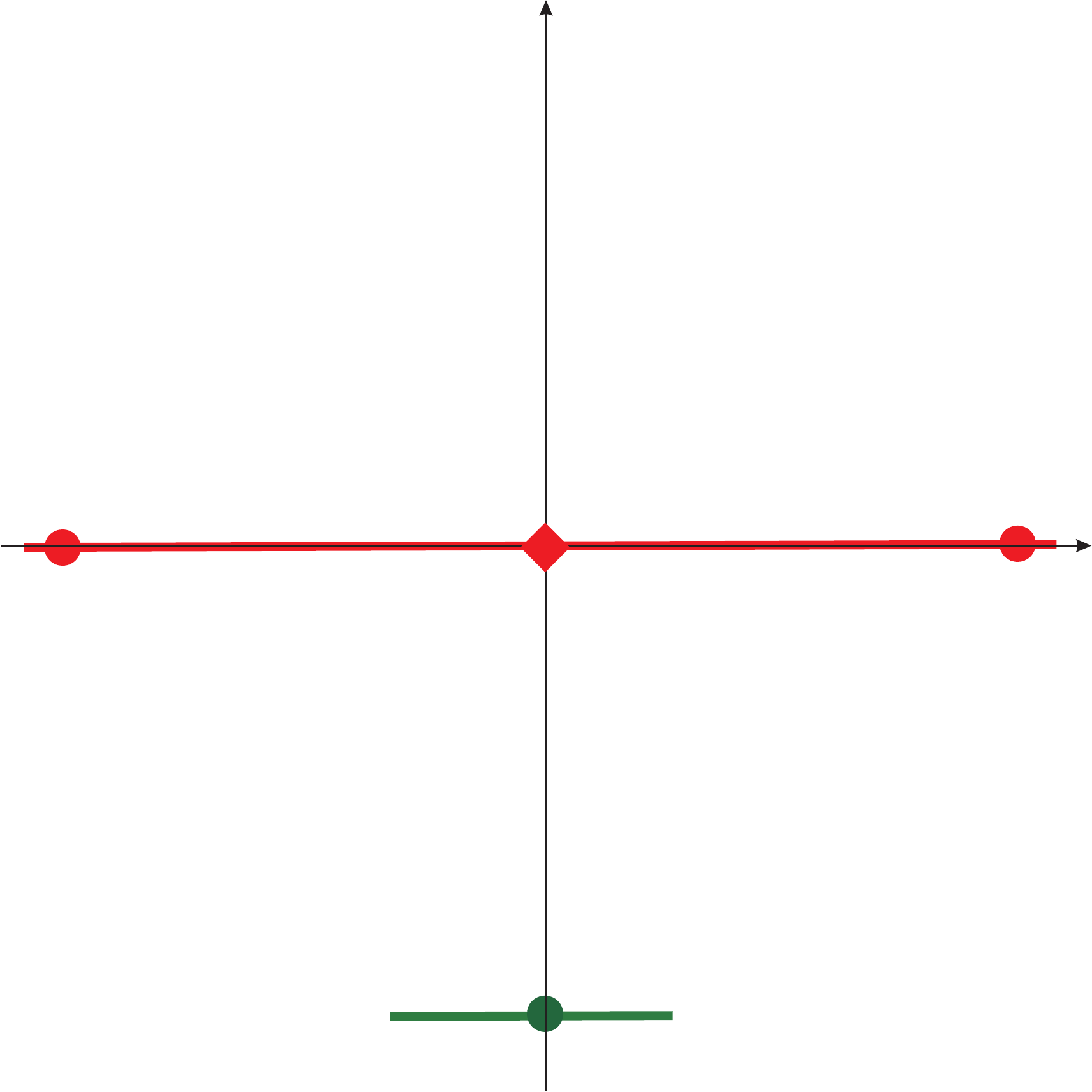}
\\
\vspace{0.3cm}
\includegraphics[width=3.5cm]{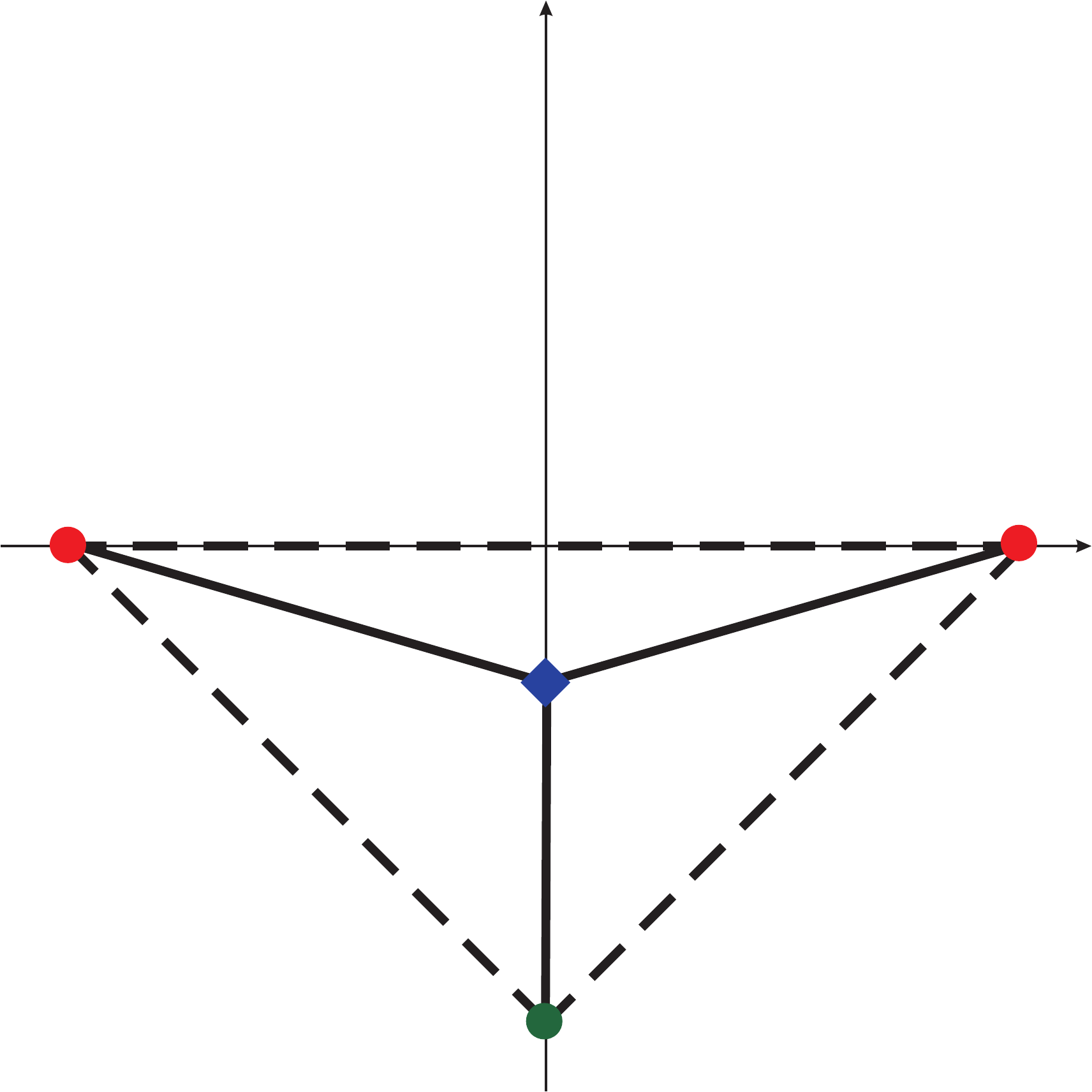}\hspace{0.3cm}\includegraphics[width=3.5cm]{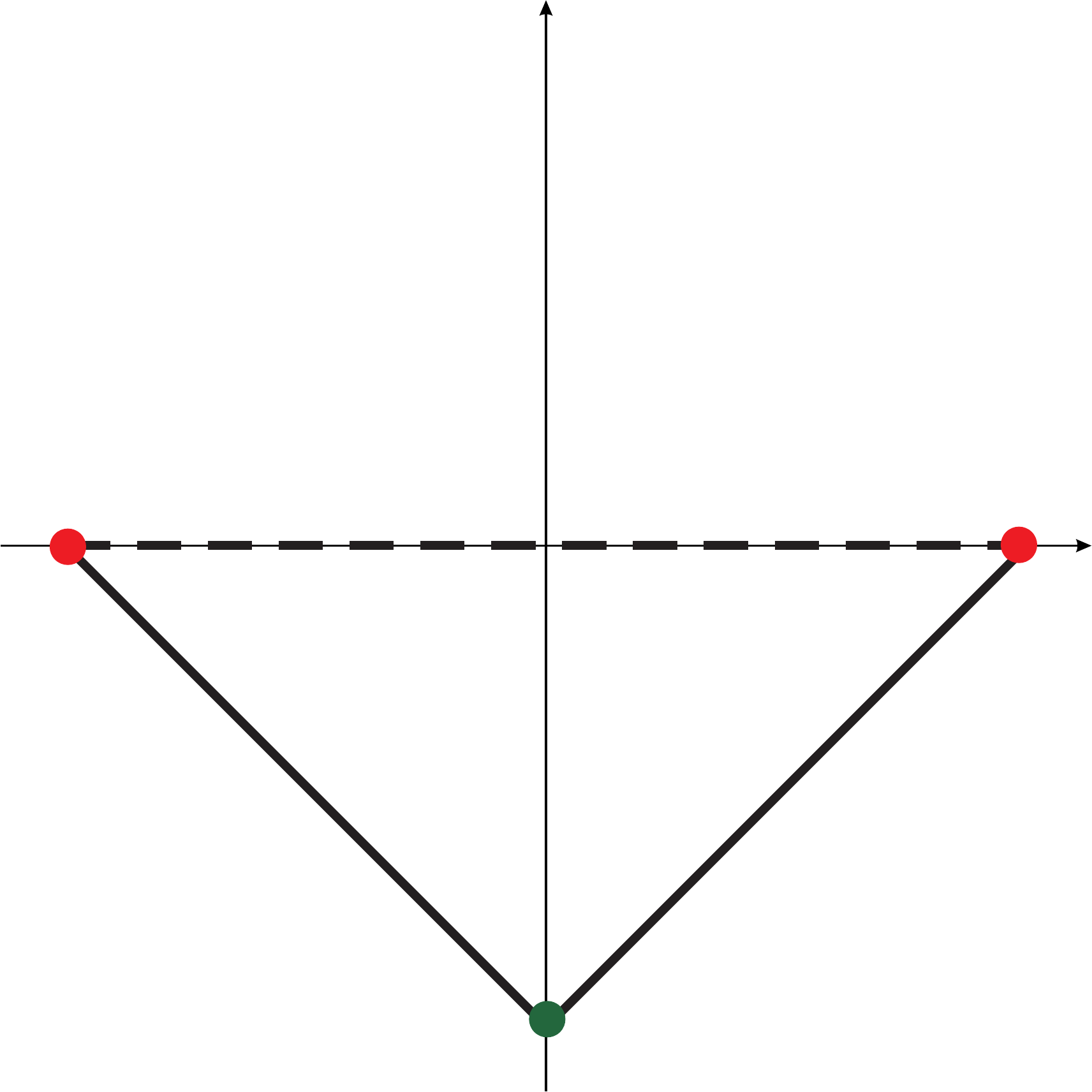}
\\
\vspace{0.3cm}
\includegraphics[width=3.5cm]{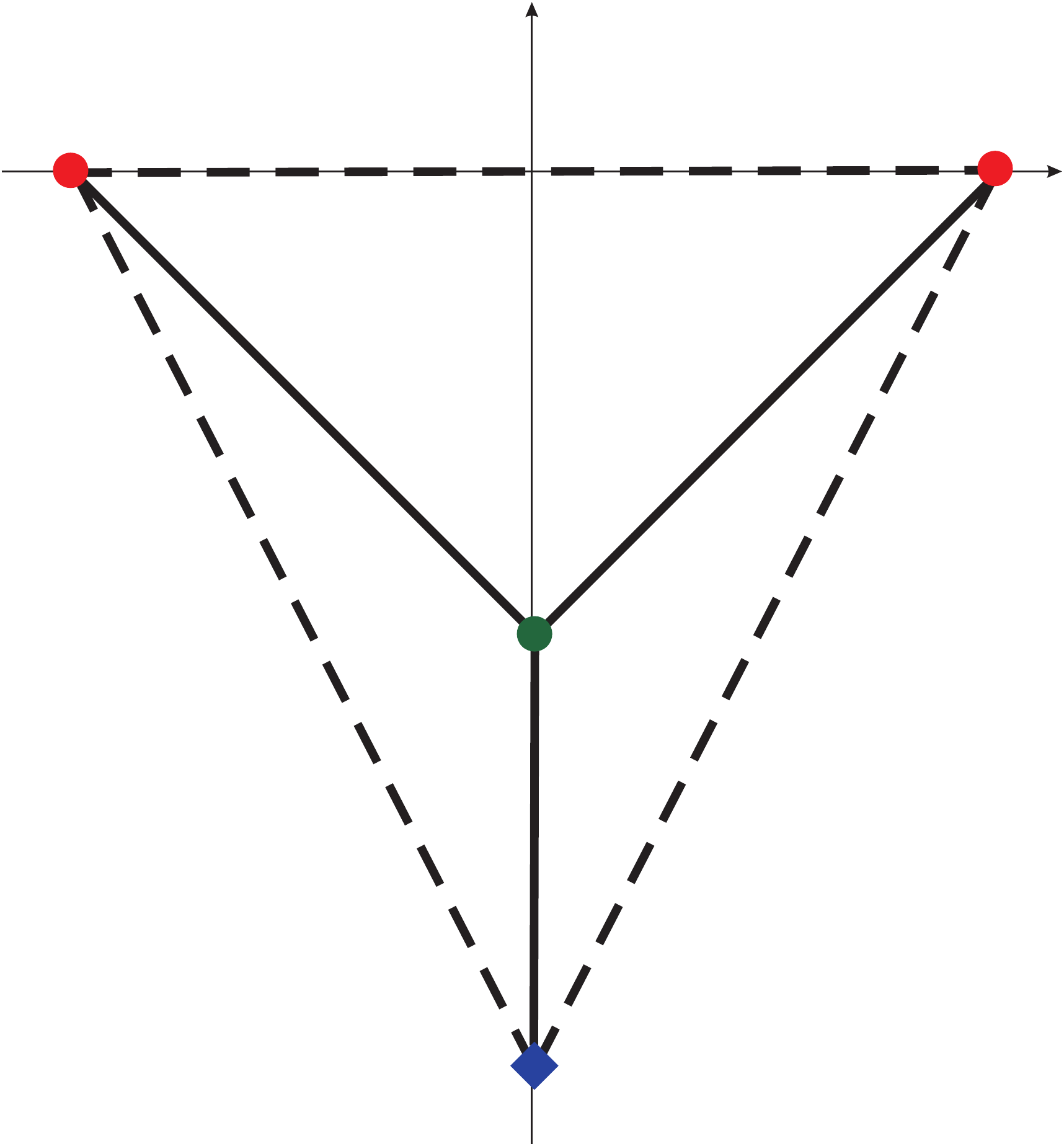}
\caption{\small{Plots of the minima of the potential. The drawings start at the left top, with $b$ given respectively by $b=1$, $b=0$, $b=-1/3$, $b=-1$ and $b=-3$.  The colors and shapes of the points are used to identify distinct values of the superpotential at the corresponding minima. Solid and dashed lines indicate BPS and non-BPS sectors, respectively. For $b=0$, the horizontal red and green lines show the continuum of minima with distinct values of the superpotential.}}
\end{figure}

In the sector $v_1\leftrightarrow v_2$ (and similarly in the sector $v_2\leftrightarrow v_3$) we can find a straight line orbit
connecting these two minima. In the sector 
$v_2\leftrightarrow v_3$ if we require that $\phi=\chi+1$ then the potential becomes (see Eqn. (\ref{potours}))
\be
V(\chi)=\frac12  \beta^2 (\chi+1)^2\chi^2\,,
\ee 
with $\beta=(5b^2/2+b+1/2)^{1/2}$.
The solution is
\be\label{slsol}
\chi(x)=-\frac12 + \frac12\tanh\left(\beta\,x\right)\,,
\ee
so that the corresponding tension of the straight line solution is
\be
\tau_{23}^{sl}=\frac{\beta}{3} \,. \label{tau23sl}
\ee

\begin{figure}[t!]
\includegraphics[width=3.5cm]{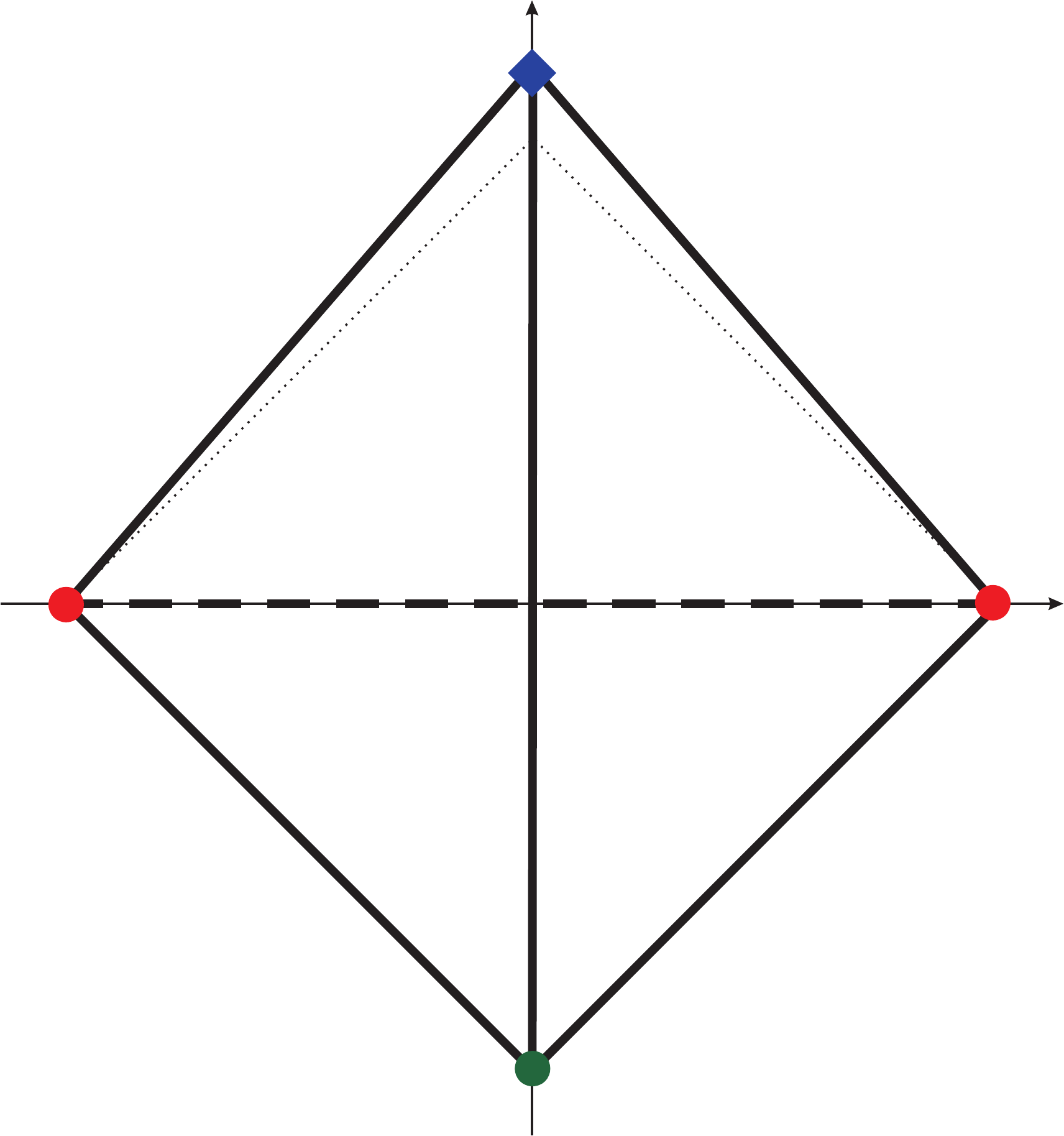}\hspace{0.3cm}\includegraphics[width=3.5cm]{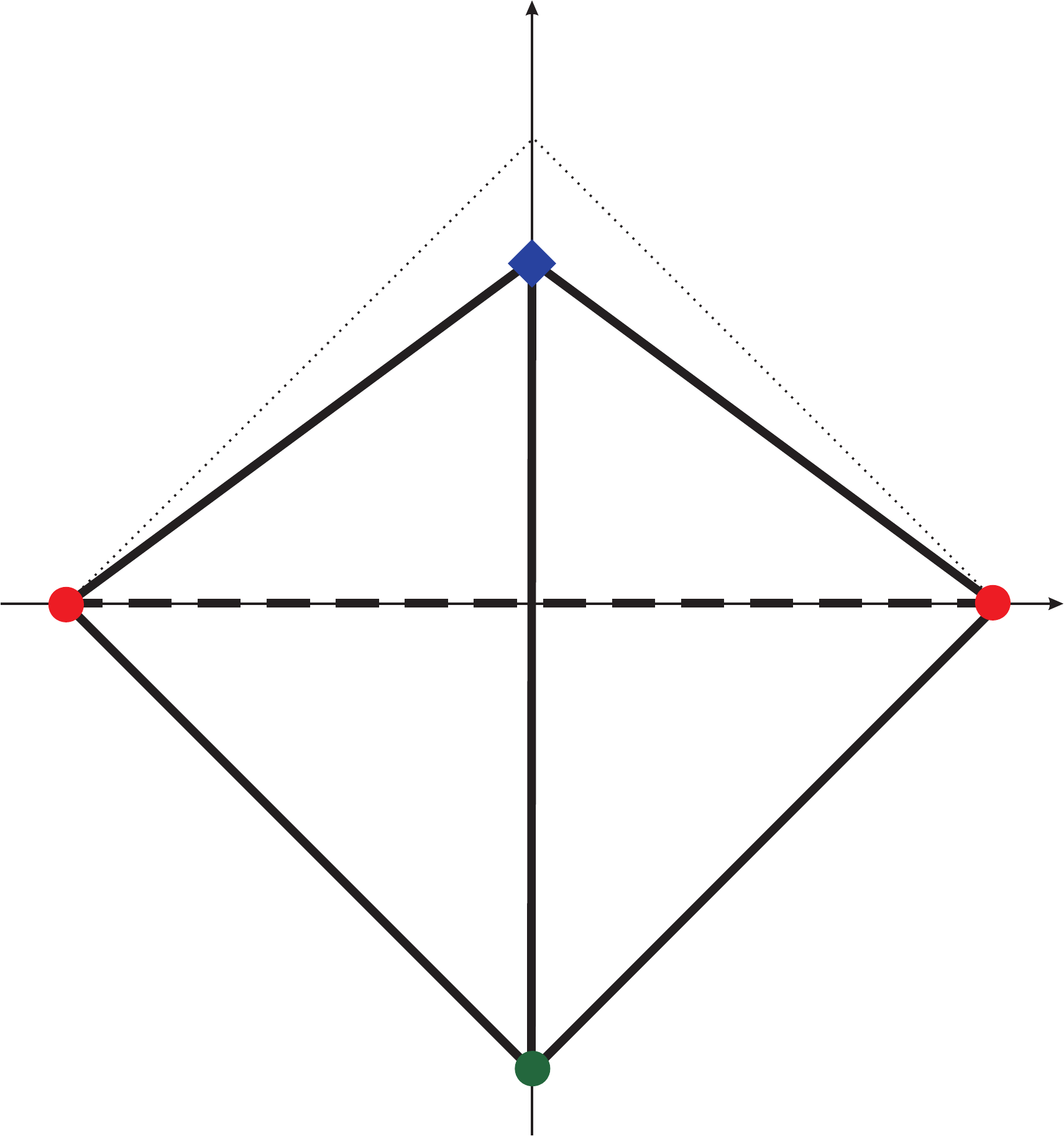}
\\
\vspace{0.3cm}
\includegraphics[width=3.5cm]{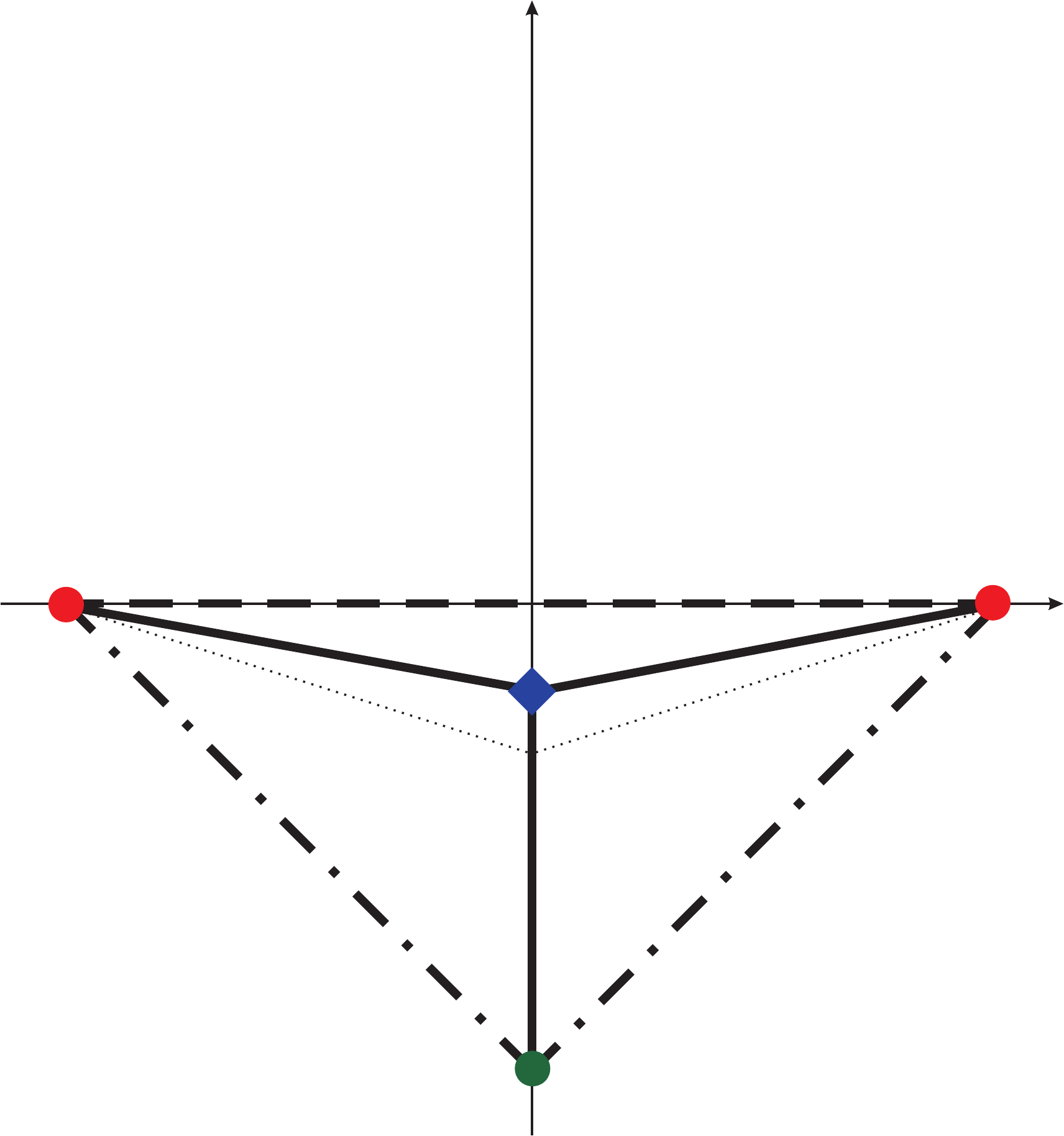}\hspace{0.3cm}\includegraphics[width=3.5cm]{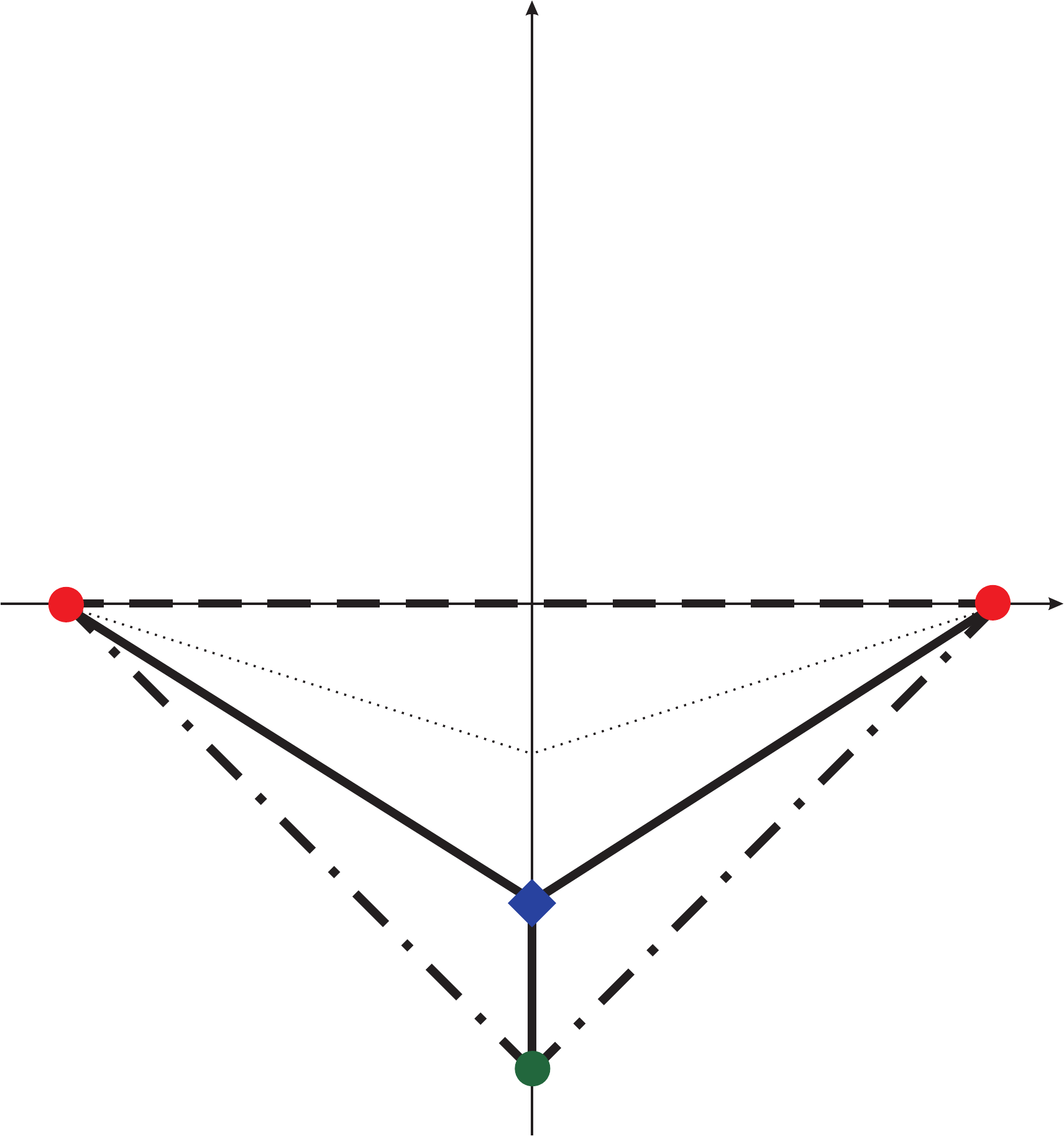}
\\
\vspace{0.3cm}
\includegraphics[width=3.5cm]{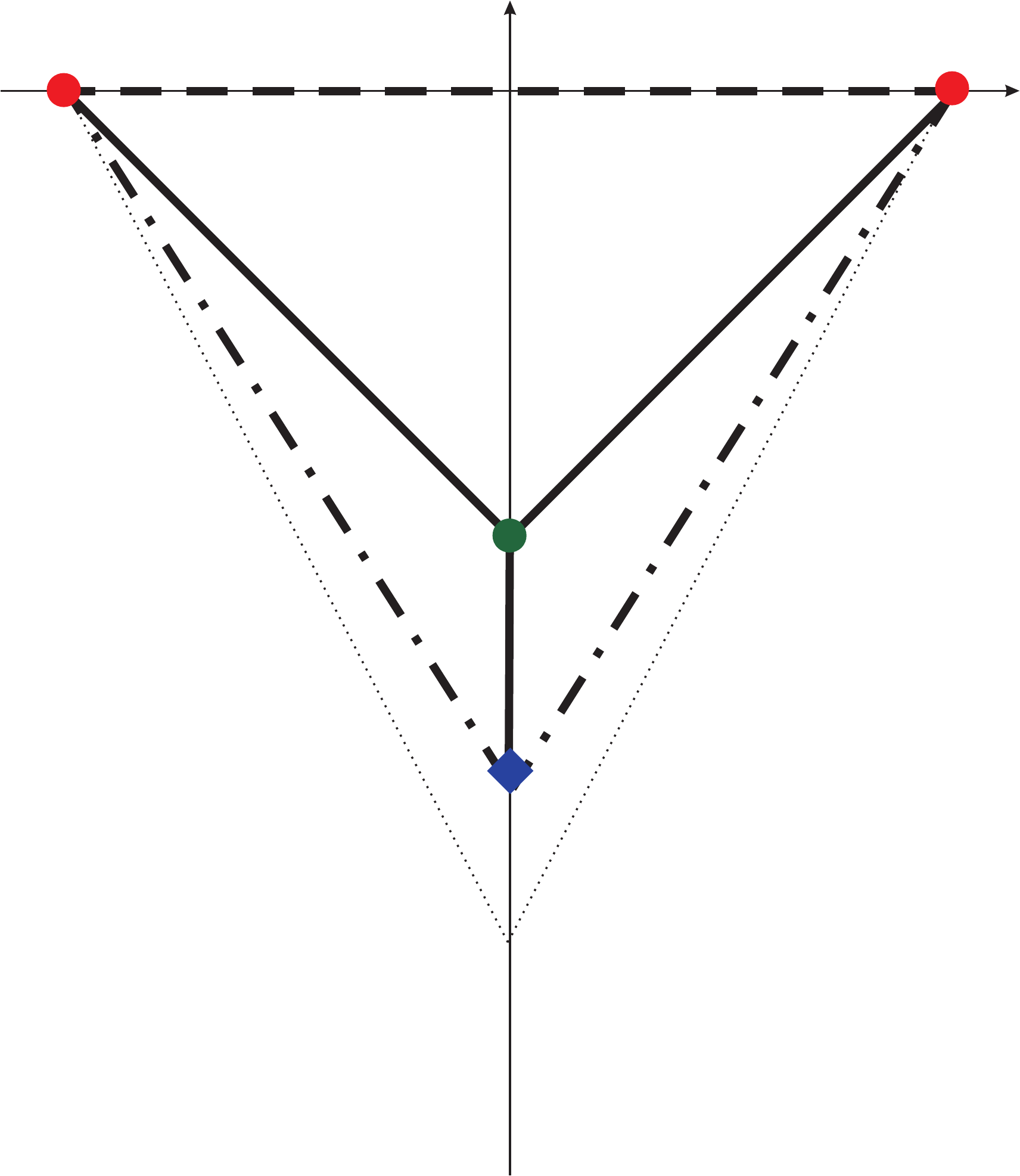}\hspace{0.3cm}\includegraphics[width=3.5cm]{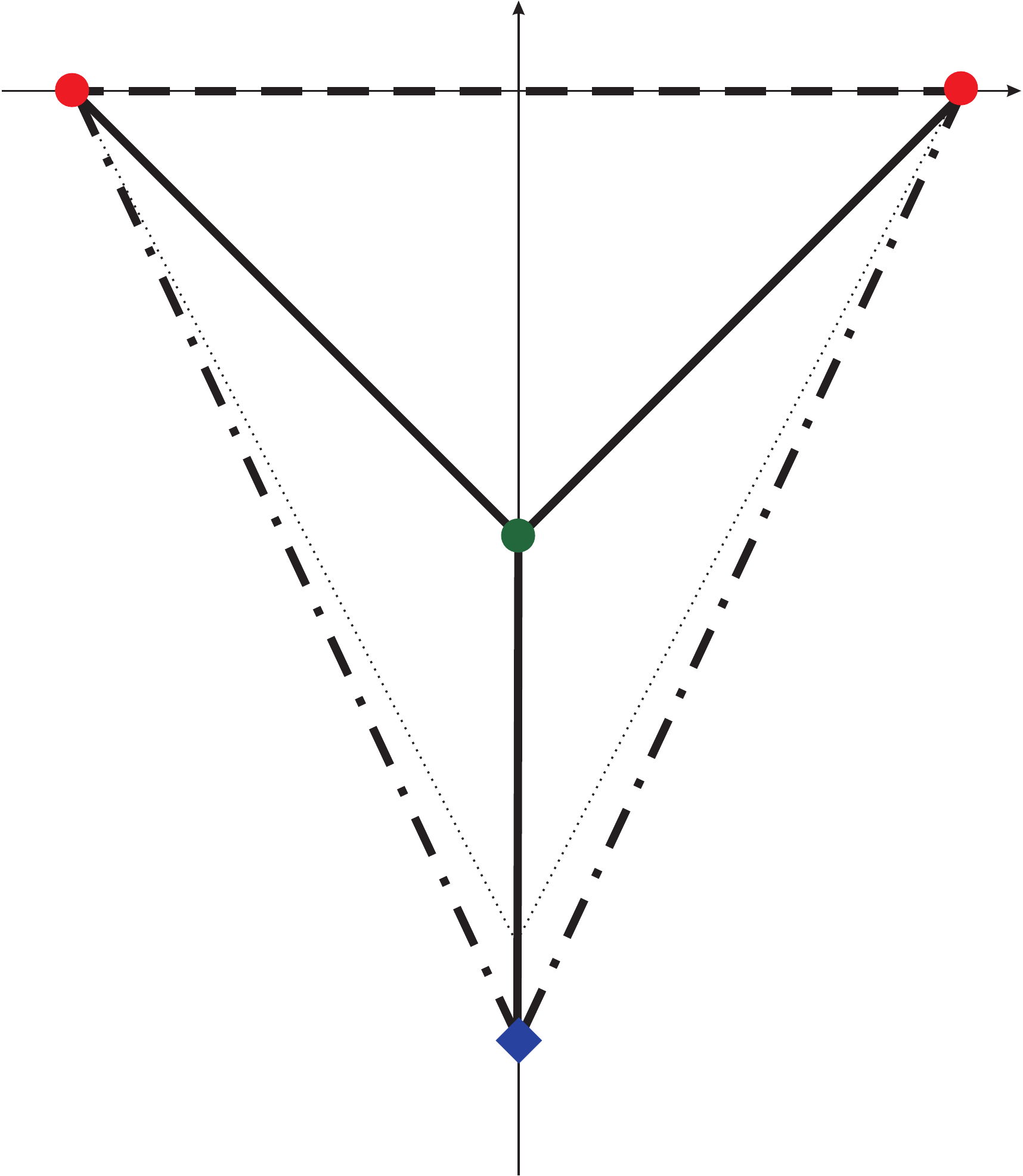}
\caption{\small{Plots of the minima of the potential. The colors and shapes of the points are used to identify distinct values of the superpotential at the corresponding minima. Solid and dashed lines indicate BPS and non-BPS sectors, respectively. The dot-dashed lines represent BPS sectors whose orbits in field space go through infinity and are never realized in practice. The drawings start at the left top, and are given for some tipical values of $b$, in the intervals $b>1$, $0<b<1$, $-1/3<b<0$, $-1<b<-1/3$, $-3<b<-1$, $b<-3$ respectively.}}
\end{figure}

There are similar expressions in the sector $v_1\leftrightarrow v_2$ ($\tau_{12}=\tau_{23}$). 
Interestingly, we note that,  except for $b=-1/3$, these two sectors are BPS sectors and they have the tension $\tau_{23} < \tau_{23}^{sl}$. We compare this with 
Eqn. (\ref{tau12}) to conclude that, in this sector, there are no straight line solutions to Eqns. (\ref{phieq}) and (\ref{chieq}), except when $b=1$. Still, Eqn. (\ref{tau23sl}) provides an upper bound to the tension of a static domain wall in the sectors $v_1\leftrightarrow v_2$ and $v_2\leftrightarrow v_3$. In the case of BPS solutions this turns out not to be very helpful since the exact values of the domain 
wall tensions can easily be calculated. However, this is no longer true for non-BPS solutions and consequently we may use the 
above procedure in order to obtain useful upper bounds on the tension of the domain walls.

\begin{figure}[t!]
\includegraphics[width=8.0cm]{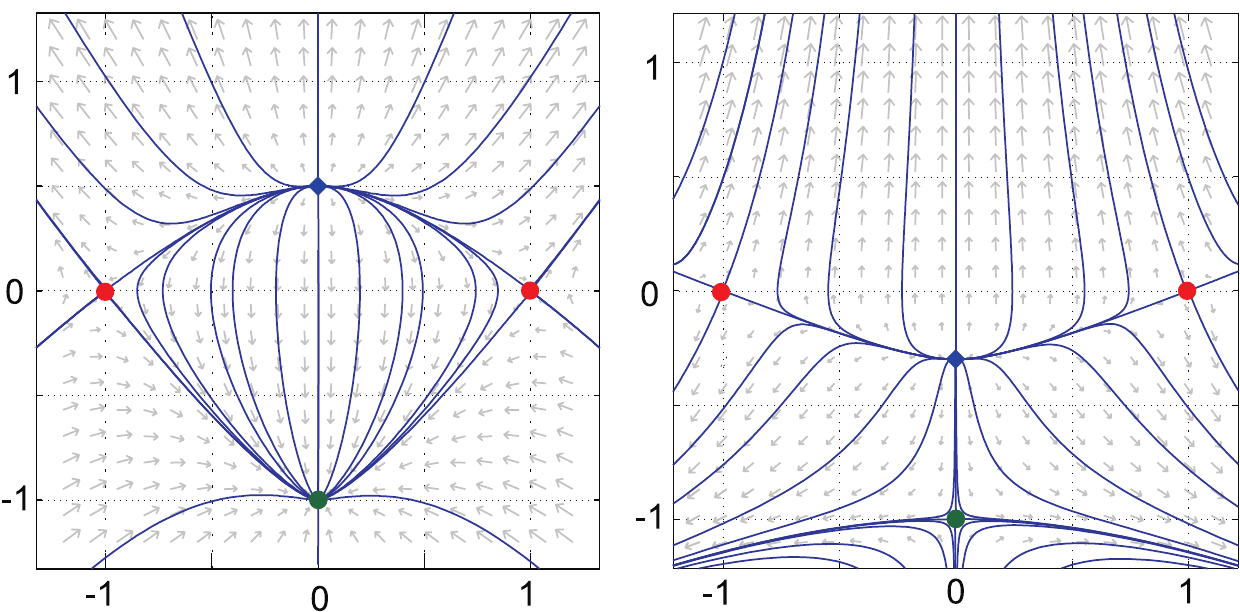}
\caption{\small{Plots of the BPS orbits for $b=0.5$ (left panel) and $b=-0.3$ (right panel). Notice that almost all the orbits are curved. The figure for $b=-0.3$ is illustrative of the fact that for  $b <0$ there are no finite paths in field space connecting the minima of the BPS sectors represented by a dot-dashed line in Fig.~5.}}
\end{figure}

For $b\neq0$ the sector $v_1\leftrightarrow v_3$ is always non-BPS. Let us find a straight line orbit connecting the two minima $v_1$ and $v_3$ for $b\neq0$. We consider $\chi=0$ to get to the potential (see Eqn. (\ref{potours}))
\be
V(\phi)=\frac12 b^2 (1-\phi^2)^2\,.
\ee
This means that the straight line non-BPS solution is
\be
\phi(x)=\tanh(bx)\,,
\ee
with tension 
\be
\tau_{13}^{sl}=4|b|/3\,.
\ee

In the lower panel of Fig.~3 we plot the values of the sum of the BPS tensions $\tau_{14}+\tau_{34}$ 
and $\tau_{12}+\tau_{23}$ for $b >0$ as well as $\tau_{13}^{sl}$, the tension corresponding to the straight line non-BPS solution.  Note that the straight line solution is expected to become increasingly accurate in the limit $b \to 1$  (so that $\tau_{13}^{sl} \sim \tau_{13}$ for $b \sim 1$).

To visualize the behavior of this family of models, in Figs.~4 and 5 we depict the minima of the potential for several values of $b$. The colors and shapes are used to identify distinct values of the superpotential at the corresponding minima. The horizontal and vertical axes represent the $\phi$ and $\chi$ fields and we indicate the BPS and non-BPS sectors with solid and dashed lines, respectively. However, these lines should not be confused with orbits in field space which in general are curved.

\begin{figure}[t!]
\hspace{-0.3cm}\includegraphics[width=4.5cm]{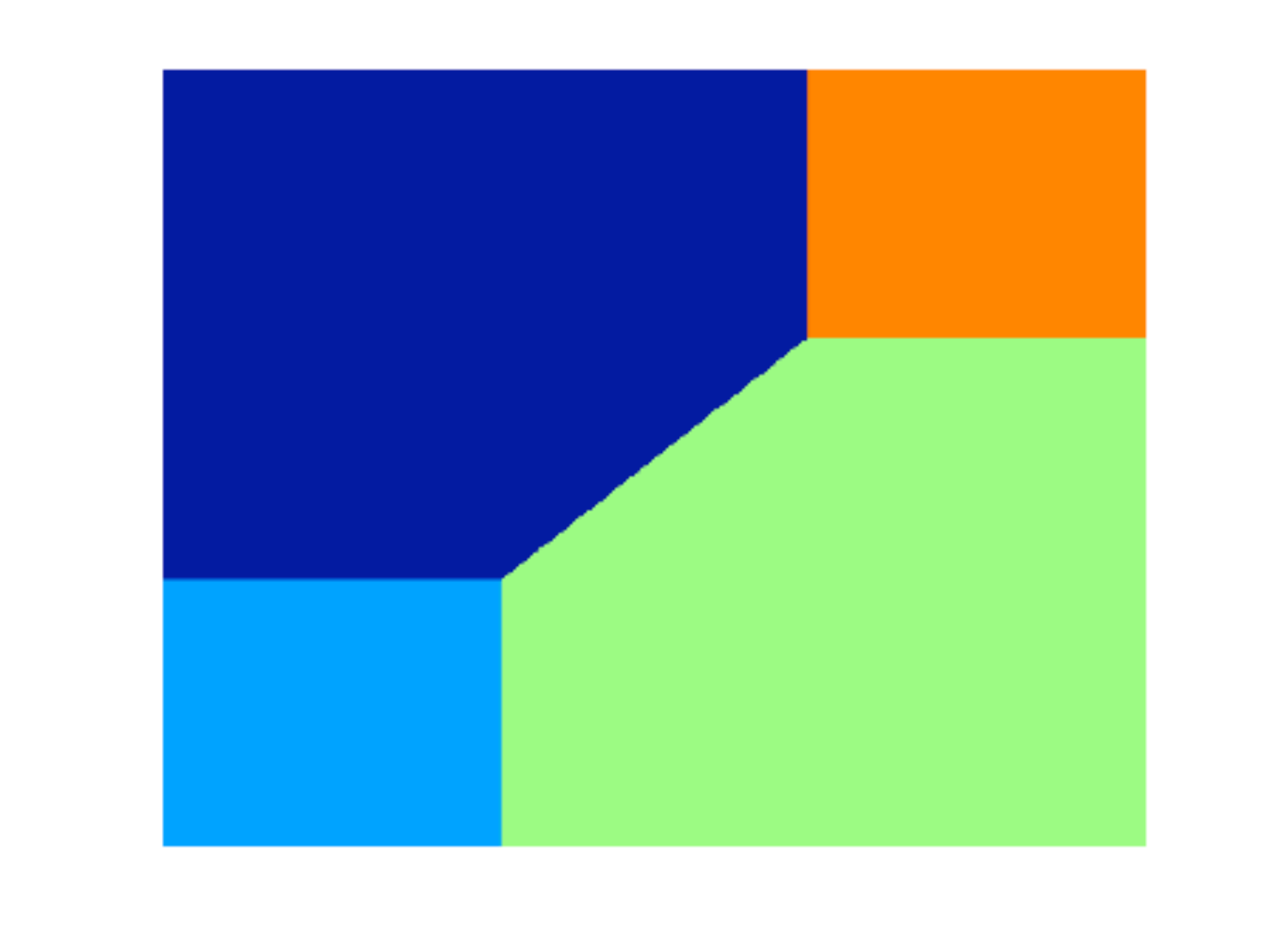}\hspace{-0.3cm}\includegraphics[width=4.5cm]{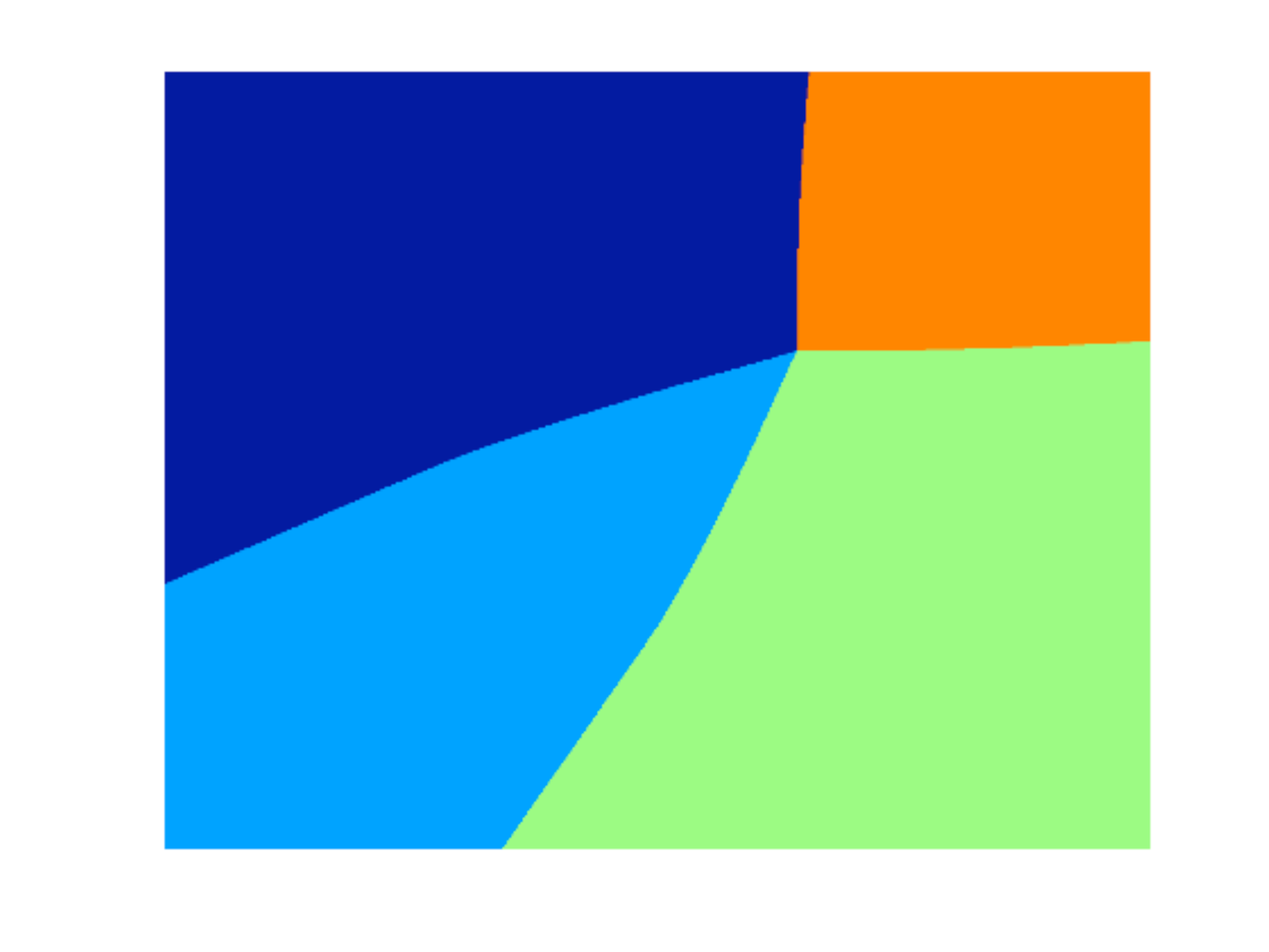}
\\
\vspace{0.3cm}
\hspace{-0.3cm}\includegraphics[width=4.5cm]{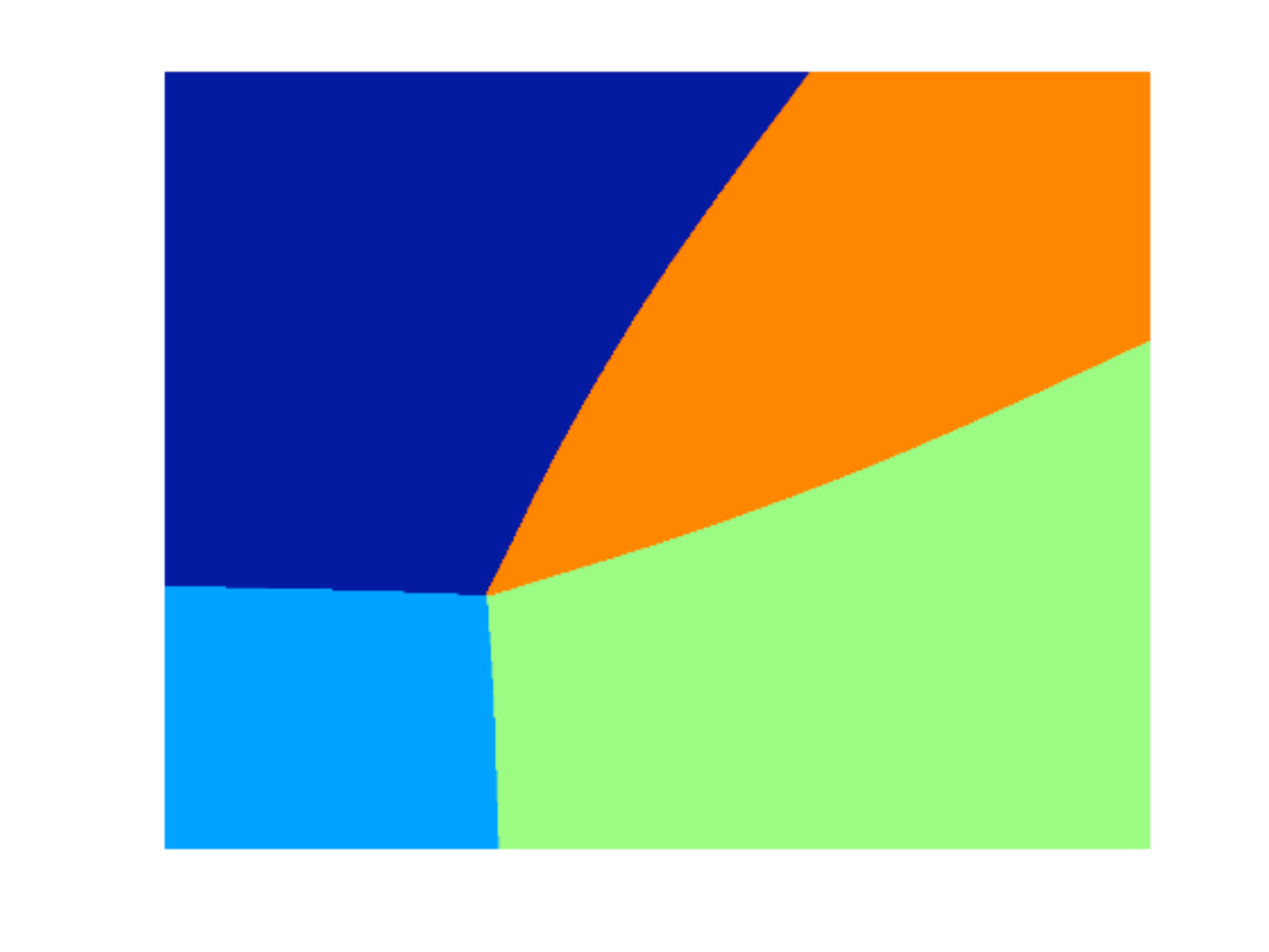}\hspace{-0.3cm}\includegraphics[width=4.5cm]{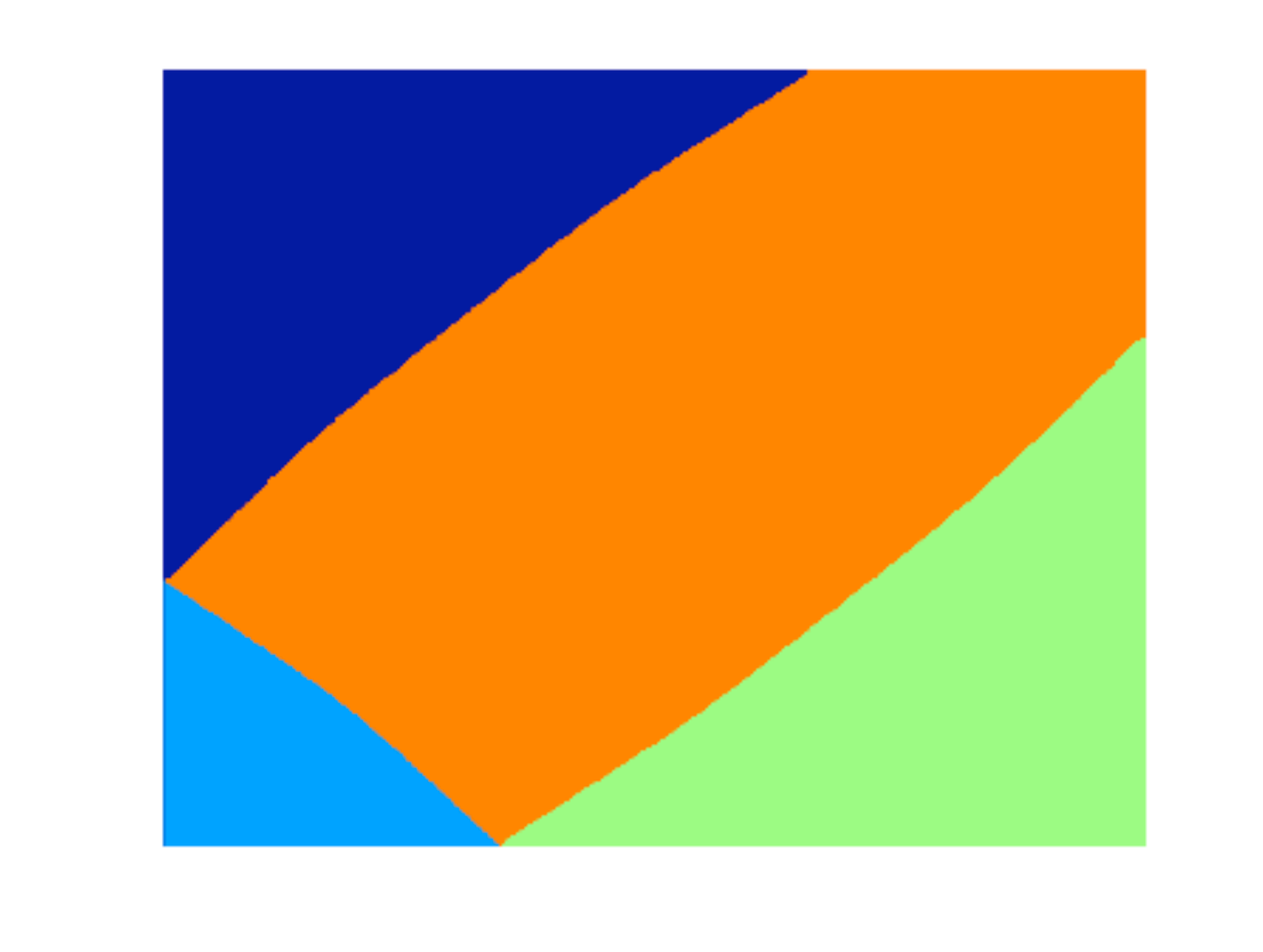}
\\
\vspace{0.3cm}
\hspace{-0.3cm}\includegraphics[width=4.5cm]{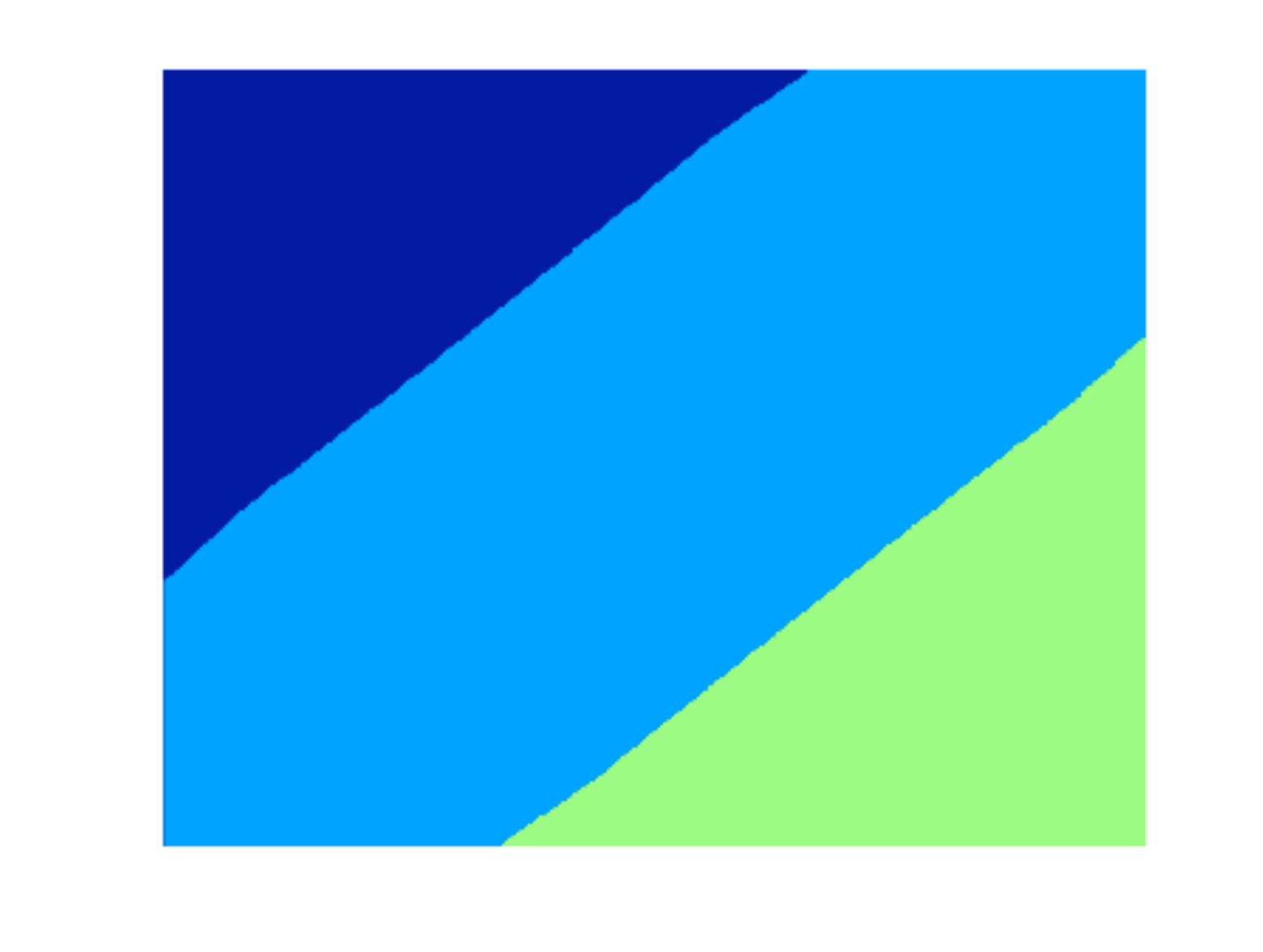}\hspace{-0.3cm}\includegraphics[width=4.5cm]{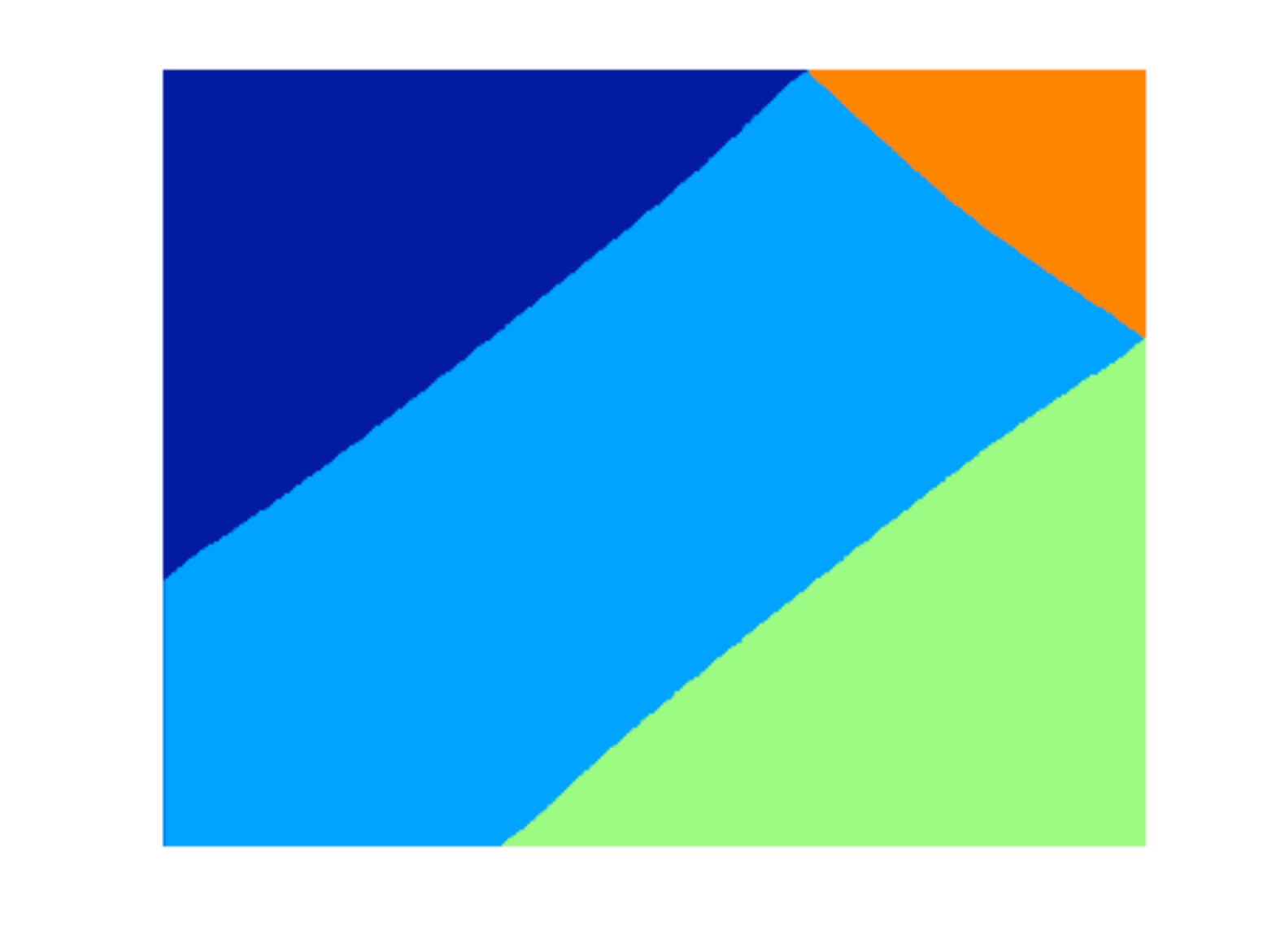}
\caption{\small{Plots of the final results of numerical simulations for $b=1.2$ (top right), $b=0.8$ (middle left), $b=-1/3$ (middle right), $b=-1$ (bottom left) and $b=-3$ (bottom right), starting from the configuration shown in the 
top left panel. The initial configuration for all these simulations is shown in the top left panel. The minima, $v_1$, $v_2$, $v_3$ and $v_4$, are represented respectively by the colors dark blue, light blue, green and orange (black, dark grey, light grey and grey in black and white).}}
\end{figure}

In Fig.~4 we represent the specific cases with $b=1$, $b=0$, $b=-1/3$, $b=-1$ and $b=-3$, which are special cases either 
because one of the BPS tensions vanishes or correspond to a case where the $Z_4$ symmetry is restored ($b=1$). We will see later that these values of $b$ correspond to critical points where the system changes behavior. For $b=0$, the horizontal red and green lines show the continuum of  minima with distinct values of the superpotential. Note that, for $b=1$, the potential is invariant 
under the transformation $(\phi,\chi) \to (\chi,\phi)$. However, the superpotential  (\ref{superpot}) is not invariant under this transformation (see top left panel of Fig.~4).

In Fig.~5 we represent some generic cases, for $b$ at some tipical values inside the several regions delimited by the values shown in Fig.~4.  The dot-dashed lines represent BPS sectors whose orbits in field space go through infinity and are never realized in practice (these orbits must pass through a point where the energy density diverges). The BPS orbits for $b = 0.5$ and 
$b = -0.3$ are given in Fig.~6. (left and right panels respectively). Notice that almost every BPS orbit is curved. For $b=0.5$, and in fact for any $b >0$, there are BPS orbits connecting every minima, except $v_1$ and $v_3$. On the other hand the right 
panel of Fig.~6 is illustrative of the fact that, for  $b <0$, there are no finite paths in field space connecting the minima of the BPS sectors represented by a dot-dashed line in Fig.~5.

\begin{figure}[t!]
\hspace{-0.3cm}\includegraphics[width=4.5cm]{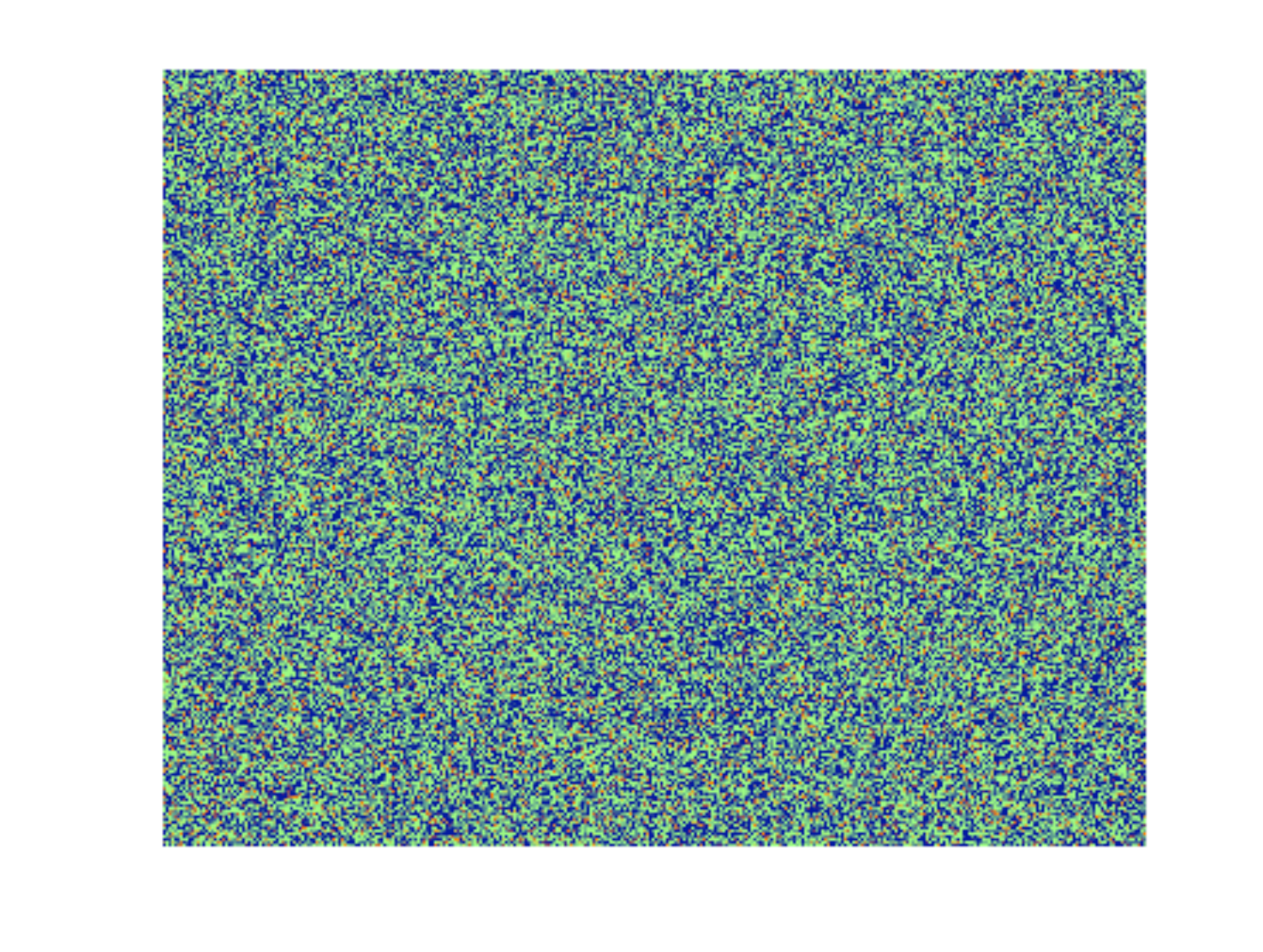}\hspace{-0.3cm}\includegraphics[width=4.5cm]{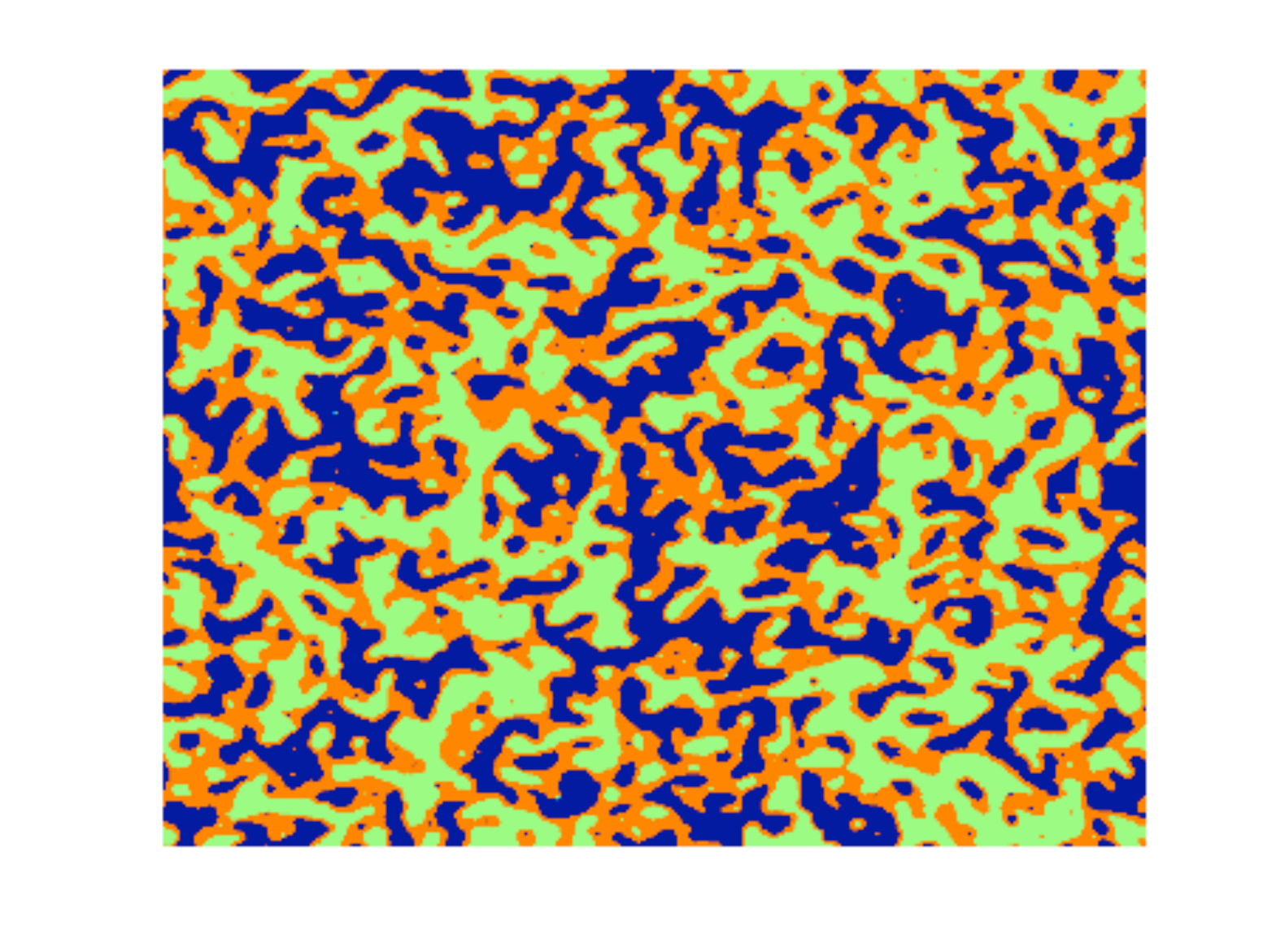}
\\
\hspace{-0.3cm}\includegraphics[width=4.5cm]{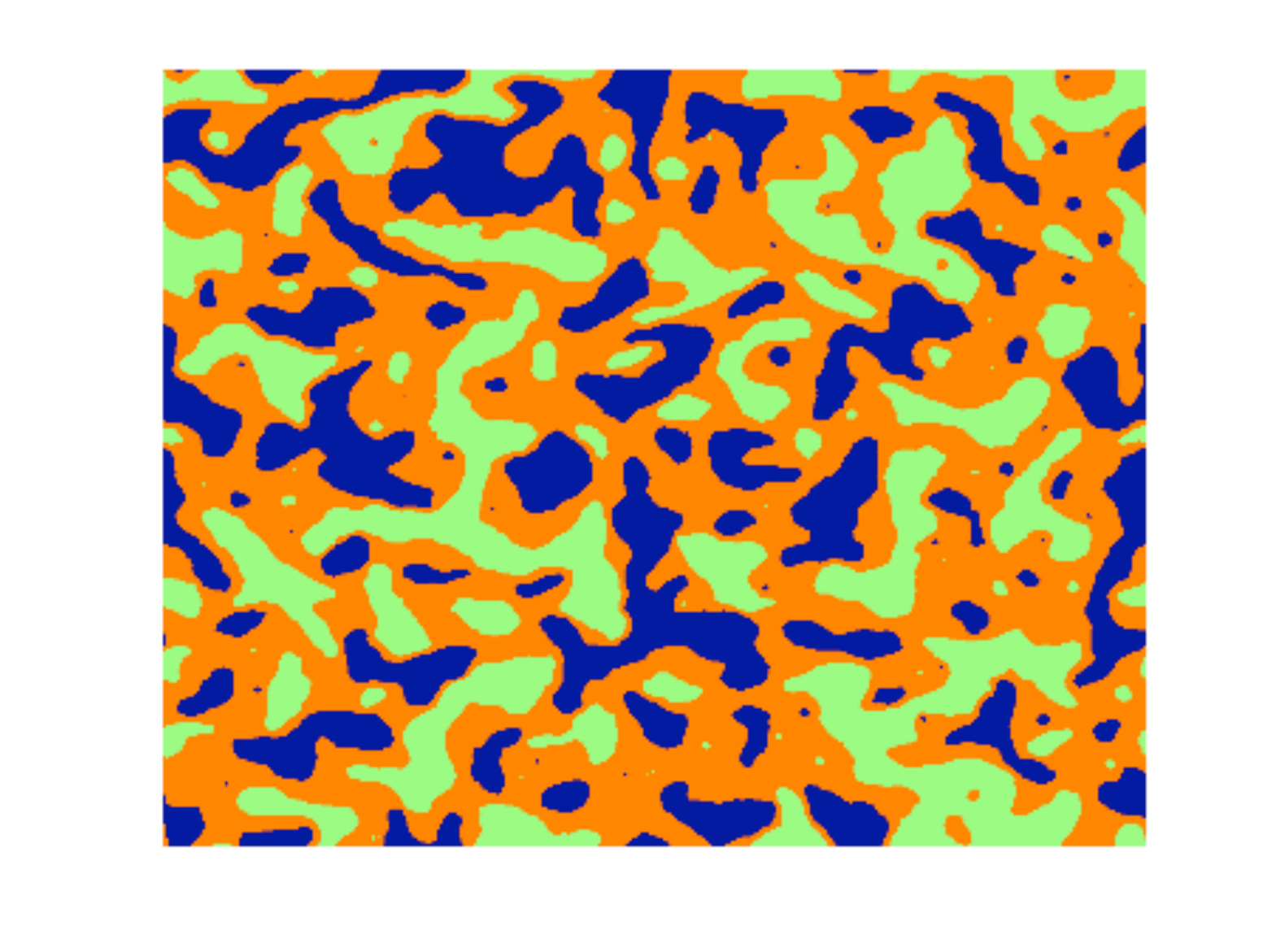}\hspace{-0.3cm}\includegraphics[width=4.5cm]{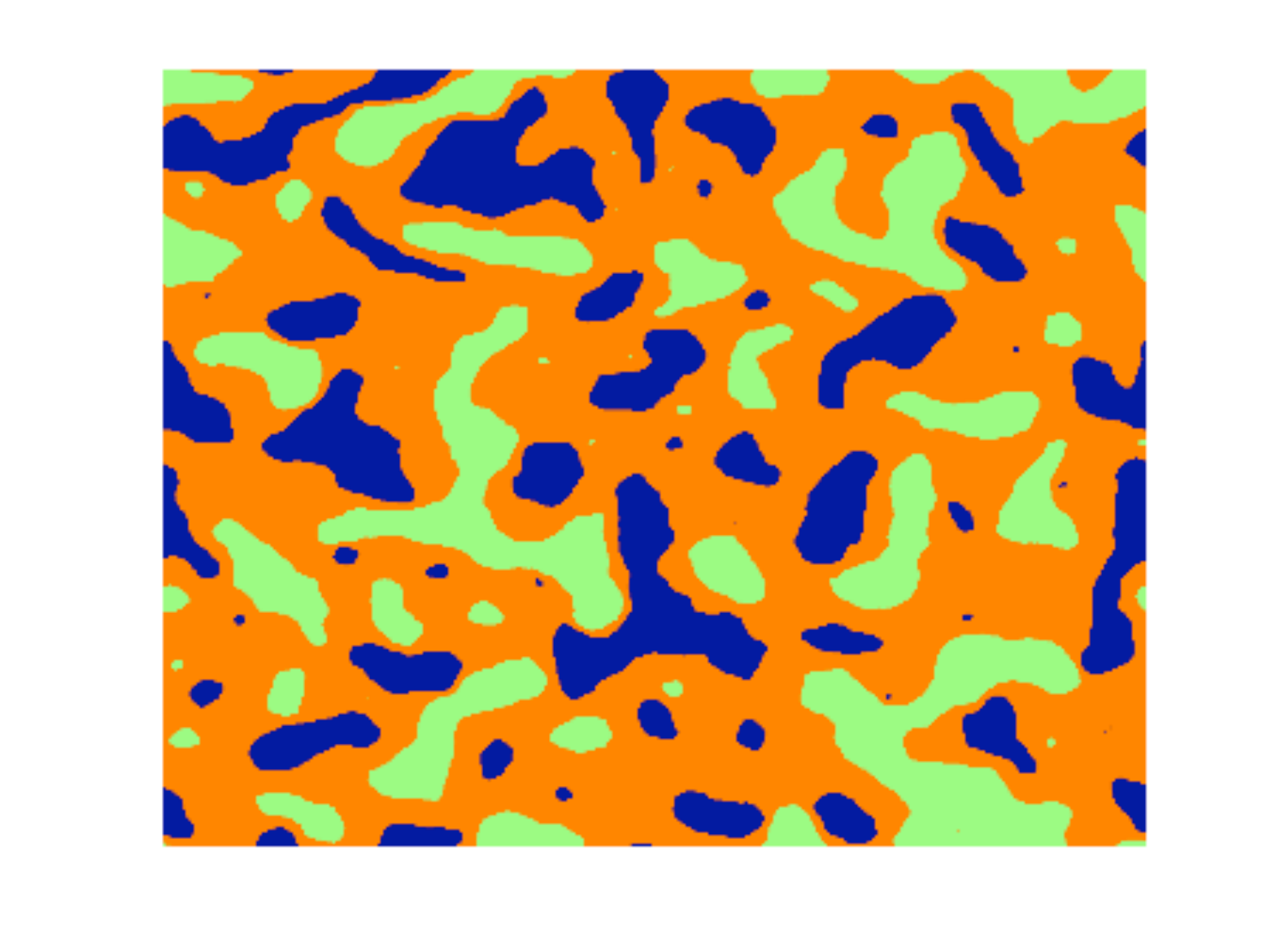}
\caption{\small{Domain wall network simulations for $b=0.98$. At the initial time the minima $v_1$ and $v_3$ were randomly assigned with equal probability to the grid points. The time derivatives of the fields, $\phi$ and $\chi$, were initially set to zero. We see that, despite not being present in the initial configuration, the minimum $v_4$ dominates at late times leading to a natural 
solution to the cosmological domain wall problem.}}
\end{figure}

We now study bifurcation and the possibility of pattern changing as we vary the parameter $b$ of our model. To do this, we numerically simulate the model for some values of $b$,  namely $b=1.2$, $b=0.8$, $b=-1/3$, $b=-1$ and $b=-3$. 
All the simulations performed in this paper are $1024^2$ two-dimensional simulations in a matter dominated universe. The 
expansion of the universe introduces a damping term which slows down the domain walls and damps the radiation produced 
by them. The initial configuration for all these simulations is shown in Fig. 7 (top left panel). The simulation 
results for $b=1.2$, $b=0.8$, $b=-1/3$, $b=-1$ and $b=-3$ are shown in Fig. 7 from left to right and top to bottom respectively, 
starting at the top right panel. Note that the results for $b=1.2$, $b=0.8$, $b=-1/3$ and $b=-3$ are representative of the behavior in the intervals $b >1$, $0 <b <1$, $-1 < b < 0$ and $b < -1$ respectively. The small curvature of the domain walls in the final 
configurations are indicative that, by the end of the simulations, the domain walls have slowed down considerably but are not yet frozen. Some very slow dynamics is still going on.

For $b>0$, $X$-type junctions are preferred.  Notice that 
$\tau_{24}=\tau_{12}+\tau_{14}$ and consequently, in one dimension, the diagonal solution in the sector  $v_2 \leftrightarrow v_4$ has exactly the same energy as the sum of the two edge solutions in the sectors  $v_1 \leftrightarrow v_2$ and $v_1 \leftrightarrow v_4$. Still, the discussion in Section II implies that $X$-type junctions are stable for any non-zero intersection angle. 
We have also numerically verified that, for any value of $b \neq 0$ the direct path in the non-BPS diagonal sector $v_1 \leftrightarrow v_3$  is always disfavored on energetic grounds. Hence, for $b>0$ only edge walls appear in the simulations. 
The intersection angles are related by $\tau_{12} \cos (\theta_4/2)=\tau_{14} \cos (\theta_2/2)$ where $\theta_2$ and $\theta_4$ 
are the angle subtended at the junction by the domains $2$ and $4$, respectively (note that $\tau_{12}=\tau_{23}$ and 
$\tau_{14}=\tau_{34}$). Consequently, the variation of the intersection angle, $\theta_{2}$, of the $X$-type junction with $b > 0$ is a consequence of the dependence of the ratio between $\tau_{14}+\tau_{34}$ and $\tau_{12}+\tau_{23}$ on the parameter $b$ 
(if $b=1$ then $\theta_2=\theta_4$). Notice (see Fig. 3, bottom panel) that the diagonal domain wall with tension $\tau_{13}$ decays into two domain walls with tensions $\tau_{14}$ and $\tau_{34}$ ($b<1$) or $\tau_{12}$ and $\tau_{23}$ ($b>1$).

For $b < 0$ there are five BPS-sectors but for two of them 
(those represented by dot-dashed lines in Fig.~5) the corresponding domain walls are never realized in practice. 
Consequently, if $b <0$ only the domain walls connecting the sectors represented by a solid line in Fig.~5 appear in the simulations, as it is clearly shown in Fig. 7.  The case with $b=-1$ (bottom left) is special in the sense that the minima $v_2$ and $v_4$ coincide.

\begin{figure}[t!]
\hspace{-0.3cm}\includegraphics[width=4.5cm]{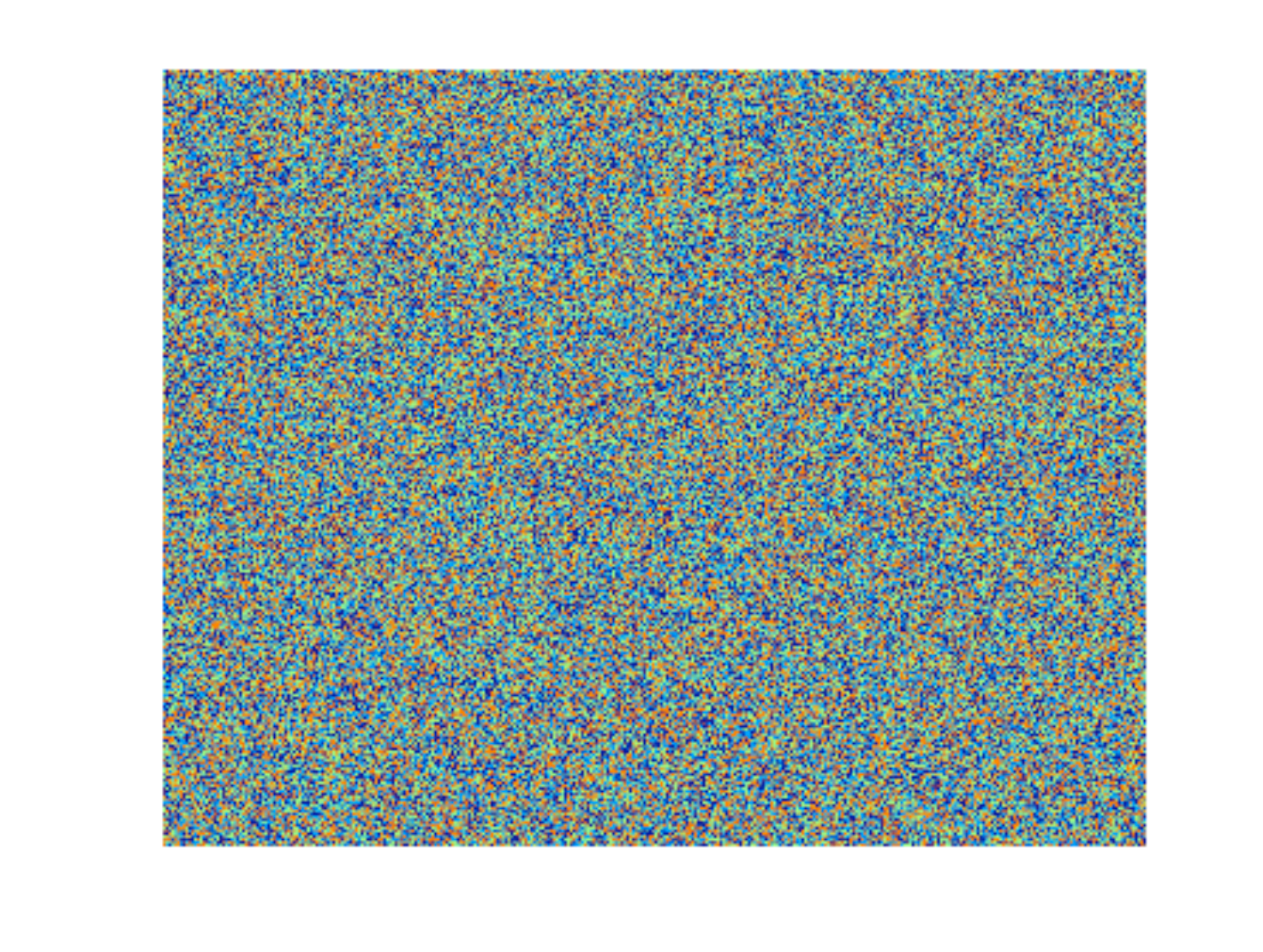}\hspace{-0.3cm}\includegraphics[width=4.5cm]{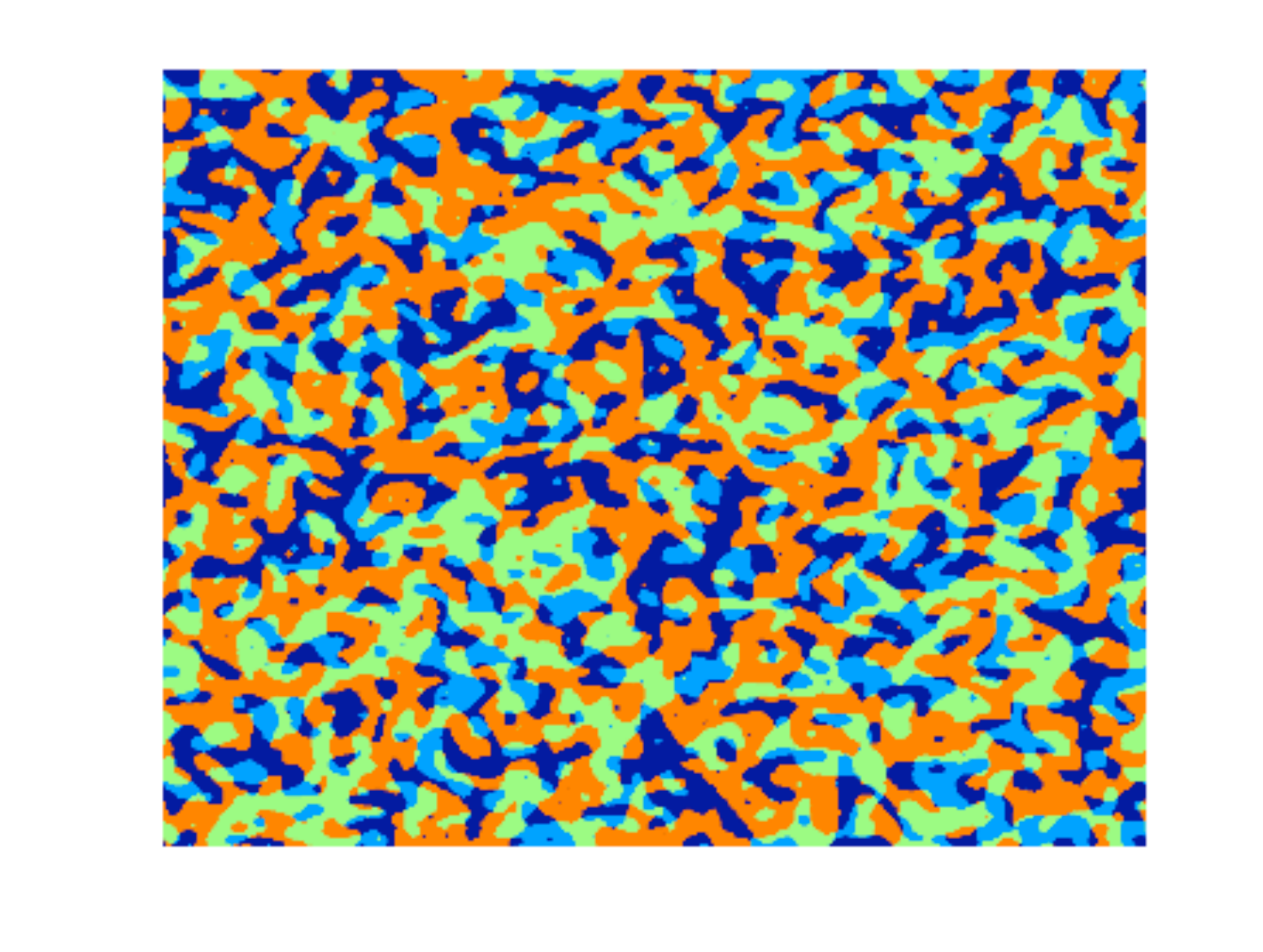}
\\
\hspace{-0.3cm}\includegraphics[width=4.5cm]{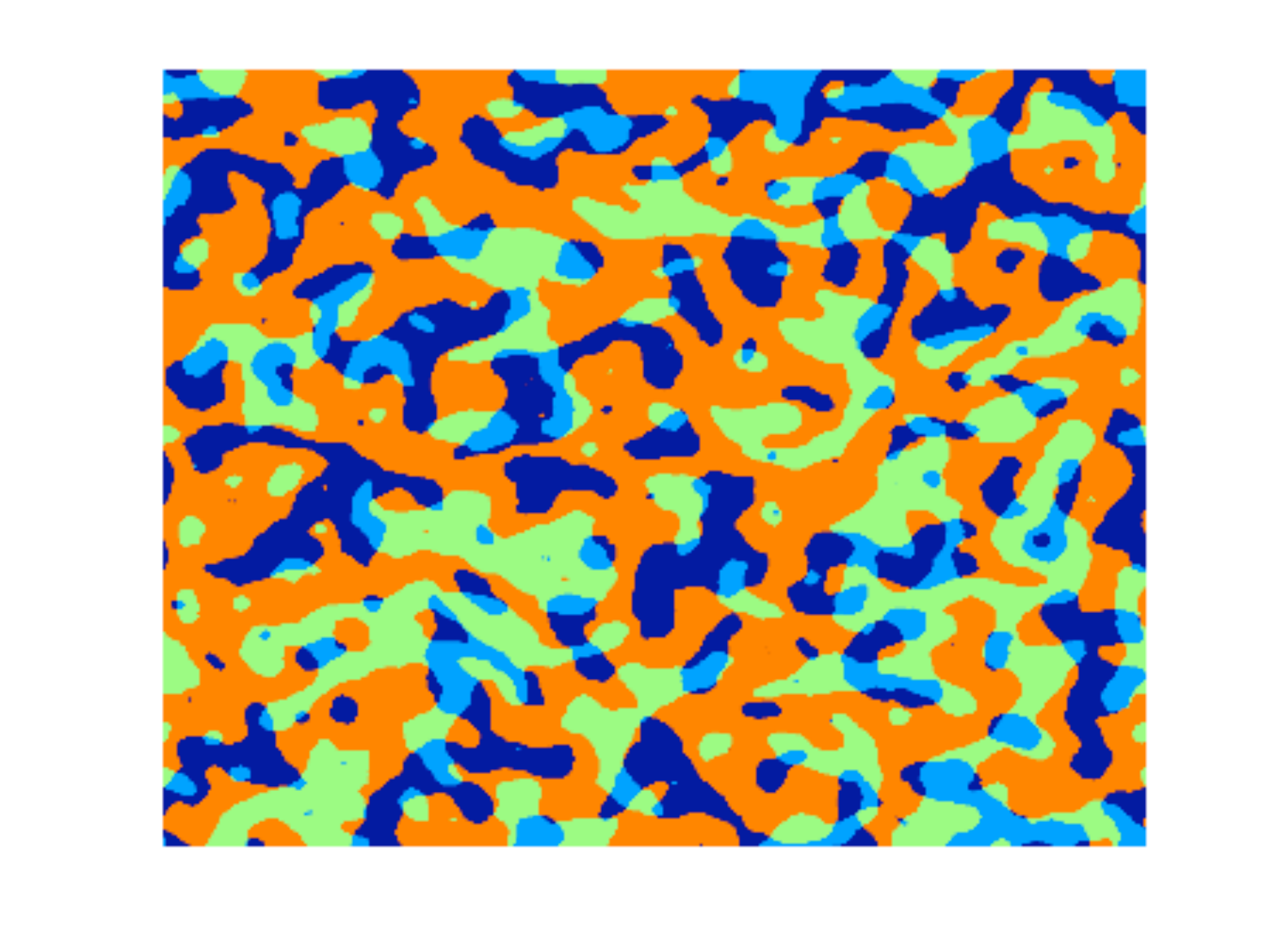}\hspace{-0.3cm}\includegraphics[width=4.5cm]{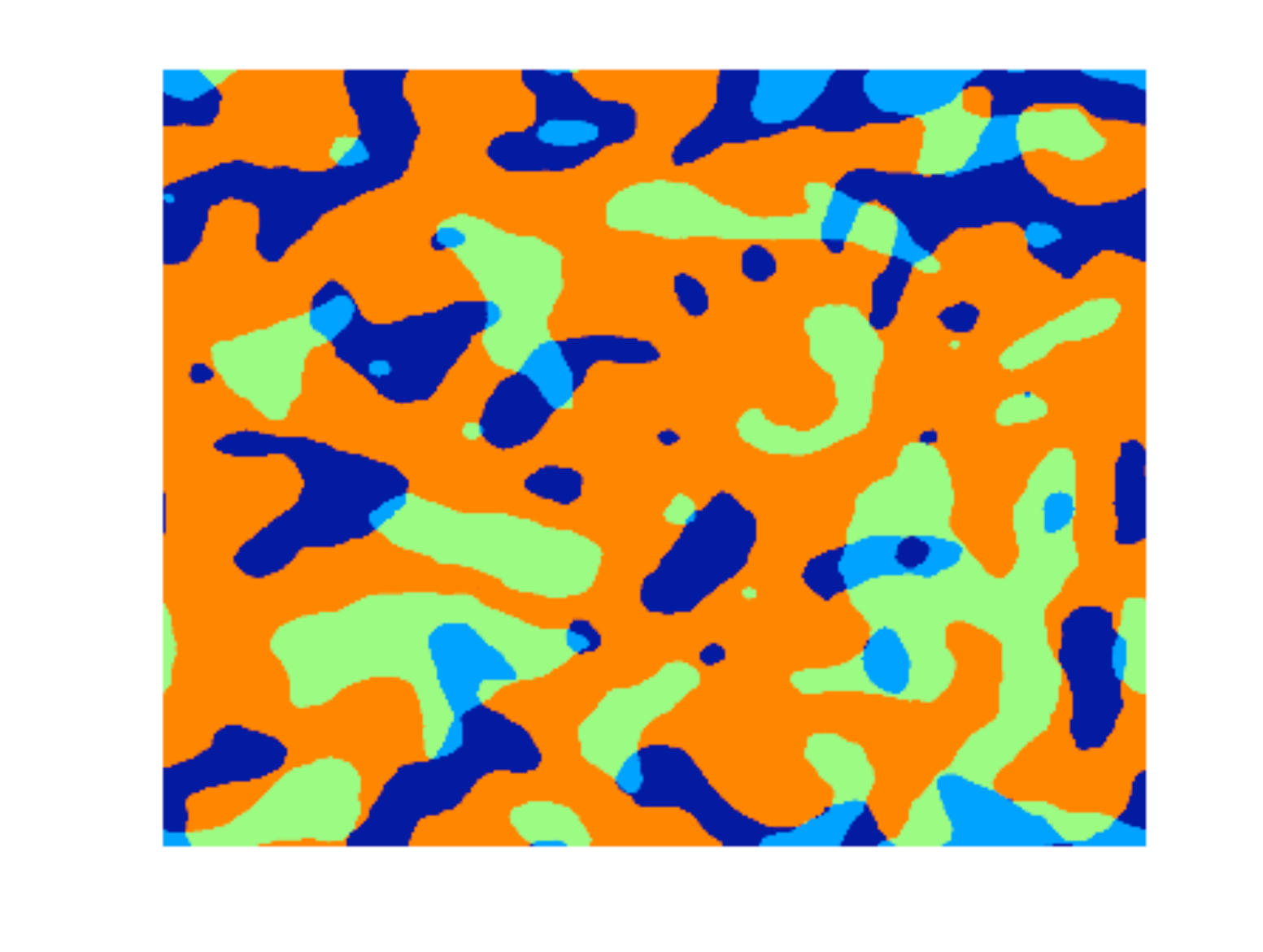}
\caption{\small{Domain wall network simulations for $b=0.98$. Initially, at each grid point, one of the minima ($v_1$ to $v_4$)  was randomly assigned with equal probability. The time derivatives of the fields, $\phi$ and $\chi$, were initially set to zero . 
In this case the junctions are of the $X$-type but the minimum $v_4$ still dominates at late times.}}
\end{figure}

\subsection{The cosmological domain wall problem} 

In the simplest scenarios domain walls are known to evolve towards a scaling regime, in which the characteristic scale of the network is of the order of the Hubble radius, until they dominate the energy density of the universe. In fact, in order to avoid catastrophic cosmological consequences, in standard domain wall scenarios, the symmetry breaking phase transition that produces the domain walls has to be smaller than about $1\, {\rm MeV}$. Biased domain wall networks may provide a way to evade the Zel'dovich bound \cite{zko} by suppressing domain walls before they have any chance of playing a cosmological role. This may be achieved if the domain walls separate regions with different values of the vaccum energy density \cite{ggk,lsw}, in which case the evolution tends to favor the lower energy minimum if the biasing is 
large enough. This is the mechanism behind the devaluation scenario \cite{fls} which has been proposed as a possible solution to the cosmological constant problem and has recently been studied in detail in ref. \cite{ams}. Another possible solution to the cosmological domain wall problem is based on initial conditions which favor one minima over the others \cite{lsw}. 

\begin{figure}[t!]
\includegraphics[width=9.0cm]{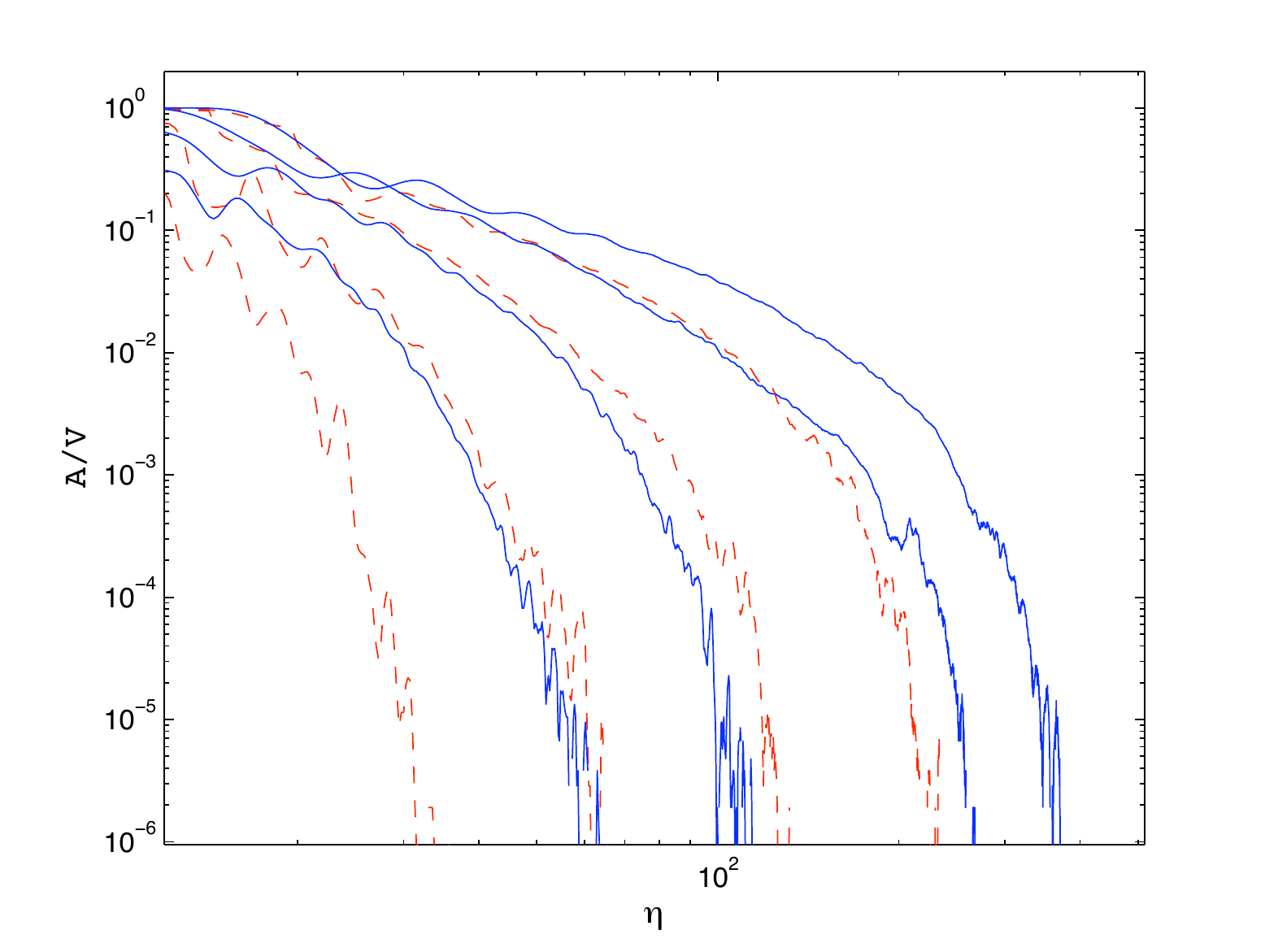}
\caption{\small{Evolution of the walls comoving area as a function of the conformal time, $\eta$, in $1024^2$ matter era 
simulations corresponding to the two situations illustrated in Figs. 8 (dashed lines) and 9 (solid lines). In this case, the asymmetry parameter is $b=0.70,0.86,0.92,0.96$ from left to right, respectively. }}
\end{figure}

Here, we explicitly show that 
small asymmetries of the vacuum potential may lead to a suppression of domain walls connecting specific minima of the scalar field potential thus leading to a biasing mechanism responsible for the suppression of the domain wall network. In Fig. 8 we plot the results of a network simulation for $b=0.98$. At the initial time the minima $v_1$ and $v_3$ were randomly assigned with equal probability to the grid points and the time derivatives of the fields, $\phi$ and $\chi$, were initially set to zero. 
The domain wall connecting the minima $v_1$ and $v_3$ is disfavored and an asymmetry is created which favors the dominance of domain $v_4$ over the others. In Fig. 9 we plot the results of a domain wall network simulation of the same model but in this case 
all the four minima were initially assigned to the grid points with the same probability. The junctions are of the $X$-type but the minimum $v_4$ still dominates at late times due to the asymmetry of the potential which leads to the different domain wall tensions. We have also verified that the results for $b=1.02$ are very similar to those obtained for $b=0.98$ except that, in this case, it is the minima $v_2$ (rather than $v_4$) which dominates over the other three. Note that, in both cases we have a multi-tension network whose evolution tends to favor domain walls with lower tensions thus  introducing  a bias which favors domains surrounded by those walls. We numerically verified that, for $b < 0$, the domain wall network has no junctions and the evolution is rather trivial, since it tends to favor the only minimum of the potential which connects all other minima with BPS domain walls. Of course, this was already expected taking into account the above discussion and the 
results shown in Fig.~7 for $b < 0$.

In Fig. 10 we plot the evolution of the comoving area of the domain walls as a function of the conformal time, $\eta$, corresponding to the $1024^2$ matter era simulations presented in Figs. 8 and 9 (dashed and solid lines respectively). However, in this case, we have considered several different possibilities for the asymmetry parameter ($b=0.70,0.86,0.92,0.96$ from left to right respectively). Fig. 10 clearly shows that the domain wall network is exponentially suppressed after some time. These results are similar to those obtained in Ref. \cite{lsw} where other biasing mechanism were considered. The total suppression of the domain walls occurs for both sets of initial conditions. However, in the case where all the minima are initially randomly assigned to the grid (solid lines), this suppression happens for slightly larger values of the conformal time than in the case where only the minima $v_{1}$ and $v_{3}$ are initially present (dashed lines). As expected, for larger values of $|b-1|$ this suppression occurs at earlier times. In particular, $b=0.96$ roughly corresponds to the smallest asymmetry of the potential for which the exponential suppression of the domain wall network still occurs on the trustfull part of the dynamical range of these simulations (where the particle horizon is smaller than half of the box size). See Refs. \cite{ammmo1} for technical details of the simulations.

\section{Conclusions}

In this work we have investigated specific models described by two real scalar fields and studied the role of bifurcation in changing domain wall network geometry and topology.  First we have considered a class of models without a superpotential controlled by a small parameter, $\varepsilon$. In this case a perturbative expansion up to first-order in $\varepsilon$ or the computation of the tensions associated with straight line orbits allows for the analytical determination of the domain wall tensions in the $|\varepsilon |\ll 1$ limit. We tested the analytical results by comparing with numerical simulations and found a very good agreement for 
$|\varepsilon | \lsim 0.2$. We also confirmed that the model bifurcates which, in more than one spatial dimension, results in a change from a phase where $Y$-type junctions are favored ($\varepsilon > 0$) to another one with only $X$-type junctions ($\varepsilon < 0$). 

Note that there are significant differences on the evolution of the macroscopic properties of domain wall networks which depend on whether or not junctions are present and, in the former case, on the dimensionality of the junctions. Although the presence of junctions does not seem to help much from the point of view of making domain walls a viable dark energy candidate \cite{ammmo1}, their impact on the dynamics of the network needs to be taken into account in any quantitative study of cosmological consequences of domain wall evolution. A good example are models in which the domain walls separate domains with different values of fundamental couplings. 

In this paper we have also considered generic models with two real scalar fields with a quartic potential governed by a cubic superpotential. In this case we have shown that the number of parameters affecting the ratios between the various domain wall tensions can be reduced to $4$. We then studied, in detail, a particular subset of models controlled by a single parameter  which, despite its simplicity, has revealed a very rich structure, with several regions of distinct behavior. We have also shown that these models incorporate a natural solution to the cosmological domain wall problem.  If the $Z_4$ symmetry 
of the scalar field potential is broken into $Z_2$, by moving $b$ away from unity, a bias is created which tends to favor one minimum of the potential over the others. This type of mechanism might operate in realistic domain wall models, leading to a suppression of the domain walls before they become cosmologically relevant.

\begin{acknowledgments}
This work is part of a collaboration between Departamento de F\'\i sica, Universidade Federal da Para\'\i ba, Brazil, and Departamento de F\'\i sica, Universidade do Porto, Portugal, supported by the CAPES-GRICES project. The authors also thank CNPQ, PRONEX-CNPq-FAPESQ,
and FCT for partial support. R.M. thanks CAPES for the fellowship BEX-0299/08-1.
\end{acknowledgments}


\end{document}